\pdfoutput=1

\documentclass[%
aip,
reprint,
 amsmath,amssymb,
]{revtex4-1}
\usepackage[hyperindex,breaklinks]{hyperref}
\usepackage{graphicx}
\usepackage{dcolumn}
\usepackage{bm}
\usepackage{bbm}
\usepackage[english]{babel}
\usepackage{xcolor}

\newcommand{\methodname}{Funnel Hopping Monte Carlo} 
\newcommand{\Methodname}{Funnel Hopping Monte Carlo}
\newcommand{\methodnameshort}{FHMC}
\definecolor{ccc}{HTML}{eb0c3b}
\newcommand{\cc}[1]{#1}

\begin{document}
\preprint{AIP/123-QED}

\title{\Methodname: An efficient method to overcome broken ergodicity}

\author{Jonas A. Finkler}%
\email{jonas.finkler@unibas.ch}
\author{Stefan Goedecker}%
\email{stefan.goedecker@unibas.ch}
\affiliation{ 
    Department of Physics, University of Basel, Klingelbergstrasse 82, CH-4056 Basel, Switzerland
}%

\date{\today}

\begin{abstract}
Monte Carlo simulations are a powerful tool to investigate the thermodynamic properties of atomic systems.
In practice however, sampling of the complete configuration space is often hindered by high energy barriers between different regions of configuration space which can make ergodic sampling completely infeasible within accessible simulation times.
Although several extensions to the conventional Monte Carlo scheme have been developed, that enable the treatment of such systems, these extensions often entail substantial computational cost or rely on the harmonic approximation.
In this work we propose an exact method called \methodname~(\methodnameshort) that is inspired by the the ideas of smart darting but is more efficient.
Gaussian mixtures are used to approximate the Boltzmann distribution around local energy minima which are then used to propose high quality Monte Carlo moves that enable the Monte Carlo simulation to directly jump between different funnels.

We demonstrate the methods performance on the example of the 38 as well as the 75 atom Lennard-Jones clusters which are well known for their double funnel energy landscapes that prevent ergodic sampling with conventional Monte Carlo simulations. 
By integrating \methodnameshort~into the parallel tempering scheme we were able to reduce the number of steps required until convergence of the simulation significantly.
\\
\\
\textit{This paper has been published in the Journal of Chemical Physics (\href{https://doi.org/10.1063/5.0004106}{https://doi.org/10.1063/5.0004106})}

\end{abstract}

\keywords{Funnel Hopping Monte Carlo, Smart Darting, Broken ergodicity, Lennard-Jones, Monte Carlo, Gaussian mixtures, RMSD}
\maketitle

\section{\label{sec:introduction} Introduction}

In the investigation of molecules and materials, the observed quantities are frequently not the result of the instantaneous state of the system but rather thermodynamic averages over an ensemble 
consisting of many configurations.
The theoretical calculation of such thermodynamic quantities is not a trivial task, because a large 
variety of different configurations as well as anharmonic effects have to be taken into account.
Analytical solution do therefore in general not exist and simulation methods are required. 
\cc{
 Some of the first computer simulations in condensed matter already addressed this kind of problems using Molecular Dynamics (MD) and this method is still nowadays widely used   
in many fields and in particular in biological simulations. 
The harmonic approximation also plays an important role as a starting point for 
numerous simulation methods such as the harmonic superposition 
approximation~\citep{stillinger_1982, calvo_equilibrium_2002}.
The applicability of this method is however very limited, because it does not account 
for anharmonic effects. 
Monte Carlo (MC) methods represent another pillar for such calculations.
}

The increasing speed of modern computers has enabled the application of MC methods to larger and more complex systems, making Monte Carlo simulations one of the most popular and widely used tools in the field of statistical mechanics. Nevertheless most Monte Carlo simulations are based on force fields since 
performing them at the more accurate density functional level wold be too expensive. 

A Monte Carlo simulation consists of a random walk over configuration space that generates random samples from a target distribution. 
These samples can then be used to calculate expectation values over the target distribution. 
Starting from an initial configuration, consecutive configurations 
are selected by repeated application of a proposal and an acceptance/rejection step.
The new proposed configuration $\mathbf{r^\prime}$ is then accepted or rejected \cc{with probability $\alpha$ according to} the Metropolis-Hastings criterion~\citep{hastings_monte_1970}, which is given below.

\begin{equation} \label{eq:metropolis}
	\cc{\alpha}(\mathbf{r} \rightarrow \mathbf{r^\prime}) = \min \left( 1, \frac{P(\mathbf{r^\prime})}{P(\mathbf{r})} \frac{g(\mathbf{r} | \mathbf{r^\prime)}}{g(\mathbf{r^\prime} | \mathbf{r})} \right)
\end{equation}

In the above equation $P(\mathbf{r})$ represents the target distribution from which one wants to generate samples, while $g(\mathbf{r^\prime} | \mathbf{r})$ represents the probability to propose a move from $\mathbf{r}$ to $\mathbf{r^\prime}$.

Throughout the whole text boldface characters will be used to represent a whole configuration with its $3N$ coordinates. ($\mathbf{r} = (\vec{r}_{1,x}, \vec{r}_{1,y}, \vec{r}_{1,z}, \vec{r}_{2,x}, \ldots, \vec{r}_{N,z})^\top$)
If the same character is used in non-boldface with a vector arrow on top the 3 coordinates of a single atom belonging to the same configuration are meant. $\vec{r}_i$ represents therefore the x,y and z coordinates of the $i$th atom of configuration $\mathbf{r}$.

In this work the distribution of interest is always the Boltzmann distribution which is given by the equation below.

\begin{equation} \label{eq:boltzmann}
	P(r) = \frac{1}{Z(T)} \exp \left( \frac{-E(\mathbf{r})}{k_B T} \right)
\end{equation}

Here $k_B$ is the Boltzmann constant, $T$ the temperature, $E(\mathbf{r})$ the energy of configuration $\mathbf{r}$, and $Z(T)$ the partition function. 

The Markovian nature of the Monte Carlo method dictates that the proposal of the next configuration must only depend on the current state of the simulation. 
Usually a new configuration is proposed by adding a small random atomic displacements to the current configuration.
In the case of rejection the old configuration has to be included into the average again. 
It is important to note here that a trade-off has to be made in the design of the proposal step.
If the displacements are chosen too big the proposed configuration will almost always be very high in energy, resulting in a small Boltzmann probability which will prevent the move from being accepted. 
If, on the other hand, the steps are chosen to be small, many of them will be accepted, but the correlation time between subsequent samples will be increased, reducing the efficiency of the simulation.
There is therefore an inherent trade-off between the size of the moves and the resulting acceptance rate.

A very useful way to characterize an energy landscape is based on its local minima. 
By assigning every configuration to the local minimum that one obtains by performing a local energy minimisation, the energy landscape can be partitioned into so-called catchment basins.
Usually these catchment basins are arranged in a cascading manner. The basins of very low energy minima are surrounded by basins with increasing energy.
In many systems several of these cascades, also called funnels, exist.
Understanding the energy landscape in therms of its funnels, basins and how they are connected can provide great insight into the dynamics of many systems~\citep{wales_decoding_2012, wales_energy_2010}.

In systems where multiple funnels are present the energy barriers between these funnels are much higher than the barriers between local minima in the same funnel. 
This can pose a great problem to Monte Carlo simulations because crossings of the high inter-funnel barriers occurs rarely and may not be observed during the available simulation time.
This is known as the problem of broken ergodicity.
Ergodicity means that the simulation must be able to reach any point of the configuration space where the probability is non zero. 
In theory this property is satisfied for the Boltzmann distribution as its probability is non zero everywhere. In practice however we are limited by computational power to some finite amount of Monte Carlo steps.


\cc{As many systems exist that have high energy barriers solving the problem of broken ergodicity has been of high interest and a plethora of different algorithms has been proposed. Some of these methods aim to overcome energy barriers by modifying the energy landscape using a biasing potential.
In umbrella sampling~\citep{umbrella-sampling} the potential is added before the simulation, while the metadynamics technique~\citep{laio2002escaping} dynamically adds biasing potentials during the simulation to avoid already visited regions of configuration space. However, both methods require  collective variables for the biasing. 
Other methods such as multicanonical sampling~\citep{multicanonical-algo} or Wang-Landau sampling~\citep{wang_efficient_2001} focus on sampling the inverse of the density of states to obtain a flat energy histogram. Multicanonical sampling suffers from the draw back, that the density of states has to be known a priori. This problem is overcome by the Wang-Landau method which computes the density of states dynamically during the simulation.
Despite its elegance and broad applicability the Wang-Landau method can be difficult to apply to systems with continuous degrees of freedom and requires knowledge about the range of accessible energies. If multiple funnels with high barriers are present, the Wang-Landau method can fail to achieve adequate transition rates between the funnels, as only local moves are used.  In particular the method was not able to converge to the correct result when applied to the 31 and 38 atom Lennard-Jones clusters and satisfactory results were only achieved after an order parameter was used to construct a two dimensional density of states~\citep{poulain2006performances}. 
Finding suitable collective variables or order parameters for these methods is often a challenging task, requiring in depth knowledge of the studied system, and might not always be possible.}

\cc{One of the most popular approaches nowadays is presumably parallel tempering.} With parallel tempering multiple simulations with different temperatures are run in parallel. Configurations are then exchanged between simulations with neighbouring temperatures. At the higher temperatures the simulations are able to cross the barriers. By the exchange step the information about configurations on the other side of the barriers is then propagated down to the low temperature simulations.
While this method will give exact results it has the obvious disadvantage of the additional computational cost for the higher temperature simulations. Because the Boltzmann distributions at the different temperatures are required to have significant overlap, the individual temperatures cannot be spaced too far apart and a large number of simulations at different temperatures can be necessary. This is especially the case when the temperature of interest is low compared to the temperature required for the crossing of the highest energy barriers.
Then a large number of replicas has to be inserted between the temperature of interest and the temperature required for the crossing of the barrier. Hence the propagation of configurations between those temperature ranges can become quite inefficient. 
Due to the large volume of available configuration space at high temperatures the rate at which a Monte Carlo simulation jumps between different funnels is further reduced with higher temperatures, diminishing the efficiency of parallel tempering simulations.
It is therefore advantageous to combine parallel tempering with other methods that enable the jumping between different funnels at lower temperatures~\citep{nigra_combining_2005, sharapov_solidsolid_2007}. 

In contrast to Monte Carlo simulations some modern optimization methods such as minima hopping~\citep{goedecker_minima_2004} are much more efficient at exploring the energy landscape as they employ mechanisms to avoid getting trapped in a single funnel.
Integrating these schemes into a sampling procedure however is not a trivial matter.
The difficulty arises from the need to preserve the detailed balance condition during the Monte Carlo simulation. This difficulty was overcome by~\citet{andricioaei_smart_2001} who came up with a clever way of using a set of local minima obtained prior to the Monte Carlo simulation to construct moves that directly connect the low energy regions around the local minima.

A different approach was taken by~\citet{sharapov_low-temperature_2007} who used a set of local minima to construct an auxiliary harmonic superposition system which was then coupled to the Monte Carlo simulation. \cc{Many more methods exist today that take advantage of information about the local minima~\citep{minmap, noe_boltzmann_2019,bogdan2006equilibrium,wales_surveying_2013,martiniani_superposition_2014}}

Although the number of local minima increases exponentially with the system size~\citep{stillinger_exponential_1999, doye_evolution_1999} so that in most cases it is not possible to obtain a complete set of all local minima, it is generally sufficient to only use a small subset of low energy local minima for these methods. 
If low energy minima from all major funnels are included the newly defined move helps the Monte Carlo simulation to cross the higher energy barriers between the funnels while the lower energy barriers in the system can be overcome by regular Monte Carlo moves.

Even for small model systems interacting with the Lennard-Jones potential, which is computationally very cheap to evaluate, the application of Monte Carlo simulations has only become tractable by using improved sampling methods~\citep{frantsuzov_size-temperature_2005}.
Although more accurate force field such as for examples machine learning force fields~\citep{behler_first_2017} are becoming available the high number of energy and force evaluations required by Monte Carlo simulations still limits their applicability.

Efficient sampling methods will therefore play an important role in enabling the application of Monte Carlo simulations to larger systems with computationally more demanding and accurate energy calculations.

In this work we propose a novel method called \methodname~(\methodnameshort) that also uses information about the energy landscape generated prior to the simulation to introduce a new kind of move into the Monte Carlo procedure.
\cc{
The way these moves are constructed is similar to the moves of the auxiliary harmonic superposition system~\citep{sharapov_low-temperature_2007}. The main difference is that \methodnameshort~uses Gaussian mixtures instead of relying on the harmonic approximation to obtain a single Gaussian. 
This allows us to overcome the extremely low acceptance rates which are due to the harmonic approximation failing to describe the Boltzmann distribution accurately. The Gaussian mixtures are directly fit to samples that are generated from the Boltzmann distribution and do therefore not rely on any approximation or assumption.
}


\section{Method}
\subsection{Smart Darting} 
The main idea behind smart darting is to construct new Monte Carlo moves using information about the local minima of the energy landscape.
In the original approach~\citep{andricioaei_smart_2001} this is achieved in the following way. 
Given a set of $M$ local minima $\{\mathbf{R}_i\}_{i=1,2,...,M}$ the so called darting vectors $\mathbf{D}_{ij}$ are defined as the pairwise differences between the $\mathbf{R}_i$.
\begin{equation}
    \mathbf{D}_{ij} = \mathbf{R}_j - \mathbf{R}_i \quad | \quad i \neq j
\end{equation}

Additionally epsilon regions are placed around each local minimum. A configuration is considered to be inside such a region if the Euclidean distance to the local minimum $\mathbf{R}_i$ is smaller than $\varepsilon$.
\begin{equation}
    \Vert \mathbf{R}_i - \mathbf{r}_i \Vert < \varepsilon
\end{equation}
$\varepsilon$ should be chosen such that none of the regions overlap.

The darting moves will then replace a certain fraction of the standard Monte Carlo moves. In each iteration of the algorithm a random number is drawn to decide which kind of move will be performed.
In the case that a darting move is chosen it will first be checked if the current configuration $\mathbf{r}$ is inside one of the epsilon regions. 
If not the move is considered rejected. 
If $\mathbf{r}$ lies inside one of the epsilon regions corresponding to minimum $\mathbf{R}_i$ a darting move will be proposed.
First a darting vector starting at minimum $i$ is chosen randomly.
A new configuration is then proposed by adding this darting vector to the current configuration.
\begin{equation}
    \mathbf{r^\prime} = \mathbf{r} + \mathbf{\cc{D}}_{ij}
\end{equation}
The final step is then the the acceptance or rejection of the proposed configuration by the standard Metropolis criterion (eq.~\ref{eq:metropolis}).

While this method of proposing darting moves works well in some cases there are several shortcomings.
The first problem arises when we try to apply the method to systems that are invariant under rotations and translations such as for example clusters. Defining the epsilon regions for these systems is not trivial as each minimum of the system does not correspond to a single point in our coordinate space but rather a hypersurface, induced by rotation and translation of the atoms.
Although the translational ambiguity can be removed by fixing the center of mass, the solution to the rotational problem is not so obvious. Things get even worse when we start considering systems that have multiple atoms of the same kind. 
Because these atoms are indistinguishable the system will be invariant under permutation of these atoms.

This means that in our current Cartesian coordinate space, there are $2 N! / h_\alpha$ hypersurfaces that correspond to the same configuration, with $N$ being the number of atoms in our system (if all of them are equivalent) and $h_\alpha$ is the point group order of the configuration.

Having whole hypersurfaces that correspond to one single configuration makes it impossible to define the epsilon regions as described above.
Also the darting vectors can only be defined using a single orientation and permutation of the system. As soon as atoms exchange places or the system rotates these darting vectors would not connect the epsilon regions any more.

Another problem with smart darting is that it uses spherical epsilon regions.
In reality the regions of low energy, and therefore high probability, which should be targeted by the darting moves, are often rather elipsoidal due to the presence of soft and hard modes.
Because the axies of these elipsoidal regions are in general not parallel, darting by addition of a darting vector will often miss the low energy regions around the local minima.
We therefore set out to develop a method that is able to directly target these high probability regions, without making any prior assumptions about their shape.

\subsection{Eckart space and the RMSD} 
To implement such a method it is necessary that we are able to identify similar configurations to decide if a configuration is inside one of the high probability regions. As systems consisting of identical atoms are invariant under rotations and translations it is not trivial to identify equivalent configurations in the $3N$ dimensional coordinate space.
To assign a set of coordinates to a given configuration $\mathbf{r}$ that is invariant under rotation and permutation we first select a reference configuration $\mathbf{R}$ which will be one of the local minima in our case. We then determine the optimal rotation $\mathcal{R}$ and permutation $\mathcal{P}$ of $\mathbf{r}$ so that the root mean squared deviation (RMSD) to $\mathbf{R}$ is minimal. The algorithm used to minimize the RMSD is described in section~\ref{sec:minRmsd}

The RMSD is defined as follows.
\begin{equation}
	\mathrm{RMSD}(\mathbf{r},\mathbf{R}) = \sqrt{\frac{\sum_{i=1}^{N} \Vert \vec{r}_i - \vec{R}_i \Vert^2}{N}}
\end{equation}

It should be noted here that superimposing the centers of mass always results in the minimal RMSD with respect to translation. We therefore assume without loss of generality that the center of mass for all configurations is set to the coordinate origin.

It can be shown that the RMSD between two configurations is minimal if the so called Eckart conditions~\citep{eckart_studies_1935} are met~\citep{kudin_eckart_2005}. 
The Eckart conditions are the following.
\begin{equation}
	\sum_{i=1}^N \vec{r}_i - \vec{R}_i = \vec{0}
\end{equation}
\begin{equation}
	\sum_{i=1}^N \vec{r}_i \times \vec{R}_i = \vec{0}
\end{equation}
We now define the displacement $\mathbf{d}$ as the difference between the aligned structure and the reference.
\begin{equation} 
	\vec{d}_i = \vec{r}_i - \vec{R}_i
\end{equation}
With these we can now write the Eckart conditions as follows.
\begin{equation} \label{eq:eck1}
	\sum_{i=1}^N \vec{d}_i = \vec{0}
\end{equation}
\begin{equation} \label{eq:eck2}
	\sum_{i=1}^N \vec{d}_i \times \vec{R}_i = \vec{0}
\end{equation}

From these six linear equations it follows that all displacement vectors $\mathbf{d}$, obtained from a minimal RMSD alignment, are orthogonal to the following six vectors.
\begin{equation*}
\mathbf{V}^1 = 
\left( \begin{array}{c}
        1 \\ 0 \\ 0 \\ 1 \\ 0 \\ 0 \\ 1 \\ \vdots
\end{array} \right)
\mathbf{V}^2 = 
\left( \begin{array}{c}
        0 \\ 1 \\ 0 \\ 0 \\ 1 \\ 0 \\ 0 \\ \vdots
\end{array} \right)
\mathbf{V}^3 = 
\left( \begin{array}{c}
        0 \\ 0 \\ 1 \\ 0 \\ 0 \\ 1 \\ 0 \\ \vdots
\end{array} \right)
\end{equation*}

\begin{equation}
\mathbf{V}^4 =
\left( \begin{array}{c}
        0 \\ \vec{R}_{1,z} \\ -\vec{R}_{1,y} \\ 0 \\ \vec{R}_{2,z} \\ -\vec{R}_{2,y} \\ 0 \\ \vdots
\end{array} \right)
\mathbf{V}^5 =
\left( \begin{array}{c}
        -\vec{R}_{1,z} \\ 0 \\ \vec{R}_{1,x} \\ -\vec{R}_{2,z} \\ 0 \\ \vec{R}_{2,x} \\ \vec{R}_{3,z} \\ \vdots
\end{array} \right)
\mathbf{V}^6 =
\left( \begin{array}{c}
        \vec{R}_{1,y} \\ -\vec{R}_{1,x} \\ 0 \\ \vec{R}_{2,y} \\ -\vec{R}_{2,x} \\ 0 \\ \vec{R}_{3,y} \\ \vdots
\end{array} \right)
\end{equation}
Here the vectors $\mathbf{V}^1$, $\mathbf{V}^2$ and $\mathbf{V}^3$ are obtained from equation~\ref{eq:eck1} and  $\mathbf{V}^4$, $\mathbf{V}^5$ and $\mathbf{V}^6$ are obtained from equation~\ref{eq:eck2}.

We now construct $3N-6$ basis vectors $\mathbf{B}^i$ which are orthogonal to each other as well as to the vectors $\mathbf{V}^j$
\begin{equation}
    \Vert \mathbf{B^i} \Vert = 1
\end{equation}
\begin{equation}
    \mathbf{B}^i \cdot \mathbf{B}^j = 0 \quad \forall \quad i\neq j
\end{equation}
\begin{equation}
    \mathbf{B}^i \cdot \mathbf{V}^j = 0 
\end{equation}

The $\mathbf{B}^i$ can be obtained using an orthogonalization algorithm such as for example the modified Gram-Schmidt process.

Using the $\mathbf{B}^i$ as a basis we can remove 6 coordinates from our displacement vectors $\mathbf{d}$.
These six coordinates become redundant because we fixed the rotation and translation of the configuration.
This allows us to assign a unique set of $3N-6$ coordinates to every configuration.

The $3N$ dimensional vector $\mathbf{d}$ is transformed to the $3N-6$ dimensional vector $\mathbf{d^\prime}$, using the basis spanned by the $\mathbf{B}^i$, as follows.
\begin{equation}
    \mathbf{d^\prime}_i = \mathbf{d} \cdot \mathbf{B}^i \quad \vert \quad i = 1\ ... \ 3N-6 
\end{equation}
Here $\mathbf{d^\prime}_i$ denotes the $i$th component of vector $\mathbf{d^\prime}$.

To obtain the original configuration $\mathbf{d^\prime}$ is simply transformed back to the $3N$ dimensional space and added to the reference configuration $\mathbf{R}$.
\begin{equation}
    \mathbf{r} = \mathbf{R} + \mathbf{d} = \mathbf{R} + \sum_{i=1}^{3N-6} \mathbf{d^\prime}_i \ \mathbf{B}^i
\end{equation}

\subsection{Minimizing the root mean squared deviation}\label{sec:minRmsd}
In systems of distinguishable atoms the RMSD can be considered a function of the rotation of the system as the optimal translation can be found trivially by superimposing the mean atom positions of the two systems.
In systems consisting of indistinguishable atoms however we are confronted with some kind of chicken and egg problem as the optimal rotation on one hand depends on the the permutation indicating which atoms from each systems we pair together while the optimal permutation on the other hand depends on the rotation.
Each problem by itself can be solved by known algorithms. To find the optimal rotation to a given permutation we can use an algorithm based on quaternions~\cc{\citep{Kearsley_1989, Coutsias_Using_2004, krasnoshchekov_determination_2014}}.
To determine the optimal permutation for a given rotation we can use the Hungarian algorithm~\citep{kuhn_hungarian_1955} \cc{or the shortest augmenting path algorithm~\citep{jonker1987shortest}}.
To solve the combined problems we use both algorithms in alternation until a converged solution is found.
\cc{ As each of the two algorithms will only decrease the RMSD, repeated application of them will lead to a local optimum of the RMSD.
To find the globally optimal RMSD we initialized the local optimization with different initial rotations. These rotations were represented by unit quaternions. 
  Because quaternions with opposite sign represent the same rotation each rotation has to be represented by a pair of points on opposite sides of a four dimensional unit hypersphere. To distribute these rotations as uniformly as possible 
over the hypersphere of all rotations, we put a charge on each point and minimized the Coulomb energy using a simple gradient descent algorithm with the additional constraints that the points be on the unit sphere.
These uniformly distributed  rotations  increase the chances of finding the globally minimal RMSD within a limited number of steps significantly.}

To test our algorithm we generated random configurations with an RMSD of 0.1 to the local minimum of the 38 atom Lennard-Jones cluster with the third lowest energy. We chose this configuration because it is the lowest local minimum, that has no rotational symmetry.
The RMSD of 0.1 was chosen because it is large enough so that the alignment is not trivial, but small enough to ensure no other permutation than the original one can result in a smaller RMSD.
In our experiments we found that by using 400 evenly spread initial rotations the globally minimal RMSD solution was found in 100\% of the 10000 test alignments performed. \cc{If we used random initial rotations instead, only 85\% of the alignments succeeded. Even when using 600 random initial rotations, the minimal RMSD solution was only found in 95\% of the attempts.}

\subsection{\methodname} 
Using the methods described in the previous section, we are able to uniquely map structures into a $3N-6$ dimensional coordinate space. This capability is the foundation of our novel algorithm called \methodname~(\methodnameshort) as it allows us to generate Monte Carlo moves that directly target regions of low energy.

By using some metric, which may be the RMSD or fingerprints~\citep{schaefer_computationally_2016}, we assign each point in configuration space to its nearest minimum. Thus each minimum is assigned a part of the configuration space. In our implementation we used fingerprints because they are computationally cheaper.
For each minimum $\mathbf{R}_i$ we will then define a probability distribution $q_i(\mathbf{r})$ which will live in the $3N-6$ dimensional fixed frame coordinate space, and sample the low energy region around this minimum.
These $q_i$ should cover the high probability regions as exhaustively as possible. This can be done for example by using the harmonic approximation which would result in an algorithm similar to the one proposed by~\citet{sharapov_low-temperature_2007} or by a Gaussian mixture as we will propose in the following section. 
It is important to note here that these distributions do not carry any physical meaning. How well these resemble the Boltzmann distribution does not influence the accuracy of the final algorithm as detailed balance is always satisfied. 
The $q_i(\mathbf{r})$'s just allow the \methodname~algorithm to propose better moves that are more likely to be accepted which results in a more efficient sampling. 

To propose a Funnel Hopping move we first determine the minimum $\mathbf{R}_i$ that is closest to the current configuration. We then randomly choose one of the other minima and draw a configuration from the corresponding $q_j$. The choice of the target minimum can be done completely random or one can include a transition matrix $T$ with $T_{ij}$ being the probability to choose minimum $j$ when the current configuration is closest to minimum $i$.
Such a transition matrix can be used for example to avoid proposing moves to minima that are too different in energy.
The proposed move is then accepted with probability $\alpha$ according to the Metropolis criterion.
\begin{widetext}
    \begin{equation}
        \alpha(\mathbf{r} \rightarrow \mathbf{r^\prime} ) 
        	= \min \left( 1, \exp\left(- \frac{E(\mathbf{r^\prime}) - E(\mathbf{r})}{k_B T} \right)
        	\frac{q_i(\mathbf{r})}{q_j(\mathbf{r}^\prime)}
        	\frac{T_{ji}}{T_{ij}}
        	\frac{h_{\alpha i}}{h_{\alpha j}} 
        	\right)
    \end{equation}
\end{widetext}

Here $h_{\alpha i}$ is the point group order of the $i$th minimum. 
If a minimum has a rotational symmetry, $h_{\alpha i}$ alignments with the same RMSD exist. 
All these alignments will result in different coordinates if transformed to the basis vectors $\mathbf{B}^i  $  while they describe exactly the same configuration.
It is therefore $h_{\alpha i}$ as likely to pick a configuration as $q_i(\mathbf{r})$ indicates, because $h_{\alpha i}$ points in the space of the hessian basis exist that correspond to that configuration and are equally likely.

The distributions $q_i$ play two important roles. First we can see from the above equation that the acceptance probability is proportional to $q_i(\mathbf{r})$ which means that the better the $q_i$ cover the high probability regions the higher is the expected acceptance rate. The other function of the $q_i$ is that they are used to generate the proposed configurations. Again one can see that if the $q_i$ cover the high probability regions well we will propose configurations with a low energy which will result in a high acceptance probability.

Although the Gaussian mixtures are usually quite localized they do in principle have infinite support. This means that it is possible that the proposed configuration $\mathbf{r^\prime}$ lies outside of the part of configuration space that is assigned to the minimum $j$.
This would result in a move where detailed balance is not satisfied, as the inverse move is not possible. Rejecting moves to configurations outside the region of configuration space assigned to minimum $j$ ensures that the detailed balance condition is met and no errors are introduced.

\subsection{Gaussian mixtures} 
A rather straight forward approach to define the $q_i$ is to use the harmonic approximation of the energy.
As the harmonic approximation is a quadratic function, the Boltzmann distribution of this energy will be a Gaussian distribution of the following form.
\begin{equation}
	q_i^{h.a.}(\mathbf{r}) = \frac{1}{\sqrt{\left( 2 \pi k_B T \right)^{3N-6} \mathrm{Det}\left(\mathrm{H}^{-1} \right)}} \, \exp \left[ - \frac{\mathbf{r}^\top \mathrm{H} \mathbf{r}}{2 k_B T} \right]
\end{equation}
In this equation $\mathrm{H}$ represents the Hessian matrix of the energy, transformed to the basis spanned by the $\mathbf{B}^i$. It should be noted that at every local minimum of the energy the Hessian matrix will have 6 eigenvalues that are zero. These corresponding eigenvectors coincide with the $\mathbf{V}^i$ defined above. The Hessian matrix is therefore not singular in the basis spanned by the $\mathbf{B}^i$.  

Although this approximation becomes exact in the limit of the temperature going to zero we found that at finite temperatures acceptance rates of our algorithm were very low using the harmonic approximation. For the 38 and 75 atom Lennard-Jones clusters the acceptances rates were around $0.2\%$ and $0.04\%$ respectively.  Similar behaviour was also observed in the context of the auxiliary harmonic superposition system~\citet{sharapov_solidsolid_2007}.
Using the harmonic approximation in \methodname~results in an algorithm which is very similar to the auxiliary harmonic superposition system. 
We therefore include calculations using the harmonic approximation in the following chapters for comparison.

To overcome the deficiencies of the harmonic approximation we propose a different approach to find suitable $q_i$ by using a mixture of Gaussians which is defined as follows.

\begin{equation}
	q^{g.m.}_i(\mathbf{r}) = \sum_{k=1}^m a_i^k \mathcal{N}_i^k(\mathbf{r}) \quad \vline \quad \sum_{k=1}^m a_i^k = 1 \textrm{ and } a_i^k \geq 0 \, \forall k
\end{equation}
Here the $\mathcal{N}_i^k$ represent normalized Gaussians defined by means $\mu_i^k$ and covariance matrices $\Sigma_i^k$.
Once the $a_i^k$ and $\mathcal{N}_i^k$ are determined we can generate samples from the Gaussian mixture by picking a random $k$ with probability $a_i^k$ and then drawing a random sample from $\mathcal{N}_i^k$.
To generate samples from $\mathcal{N}_i^k$ we first generate a set of random numbers drawn from a standard-normal distribution using the Box-M\"uller algorithm. We then use the Cholesky decomposition of $\Sigma_i^k$ as well as $\mu_i^k$ to transform the random numbers to the desired distribution~\citep{gentle_computational_2009}.

The parameters $a_i^k$, $\mu_i^k$ and $\Sigma_i^k$ are determined by fitting the Gaussian mixture to samples drawn from the Boltzmann distribution using the expectation-maximization (EM) algorithm~\citep{dempster_maximum_1977, bilmes_1998, gupta_2010}. 

\cc{If only a single Gaussian is fit, the resulting method is equivalent to the principle mode analysis method~\citep{brooks1995harmonic}.
Alternatively  the self-consistent phonon method~\citep{koehler1966theory, gillis1968properties} could also be used to fit a single Gaussian distribution~\citep{georgescu2012self}.
As in our method the required number of energy and force evaluations was dominated by the Monte Carlo sampling and not by the construction of the Gaussian mixtures, an improved efficiency 
in this part is not be of great advantage. What matters is the improved acceptance rates that 
can be achieved with fits that use multiple Gaussians.}

We also developed a modified version of the EM algorithm which takes advantage of the high symmetry present in many low energy configurations.
\cc{By constraining the Gaussian mixture to have the same symmetry as the local minima the number of free parameters can be reduced which leads to a better quality of the fit.}
An outline of the modified algorithm is given in appendix~\ref{sec:sym-em}.

\cc{For each of the local minima that are included, samples are generated and the Gaussian mixtures are fit individually.} The samples are collected from a standard Monte Carlo run initialized at the local minimum $\mathbf{R}_i$ after a short equilibration phase. During the Monte Carlo run we repeatedly check if the current configuration is still inside the region assigned to the local minimum $\mathbf{R}_i$. If the region was left the simulation is reinitialized at $R_i$ and some equilibration steps are performed. This ensures that the samples are all drawn from a single peak in the Boltzmann distribution that belongs to the corresponding minimum.

\cc{For our Lennard-Jones clusters we used a local geometry optimization to check if the Monte Carlo walker has left the catchment basin of the local minimum. In our experiments this required approximately 100 energy and force evaluations per sample that we collected. 
For the Lennard-Jones clusters these energy and force evaluations are extremely cheap. If a more expensive method, such as for instance DFT, would be used one would have to resort to fingerprints or RMSD calculations to check if the catchment basin has been left. We therefore decided not to include these energy and force evaluations into the final results given below. }

\section{Application}
We tested our algorithm on clusters consisting of 38 ($LJ_{38}$) and 75 ($LJ_{75}$) atoms interacting with the Lennard-Jones potential which is given below. 

\begin{equation}
	E_{LJ} = \sum_{i<j}^N 4\epsilon \left[\left(\frac{\sigma}{r_{ij}}\right)^{12} - \left(\frac{\sigma}{r_{ij}}\right)^6\right]
\end{equation}
Here $r_{ij}$ are the pairwise distances between atoms $i$ and $j$.
During each step of the simulation we decided randomly with a ten percent probability to perform a Funnel Hopping move.
All other moves were performed using a Hamiltonian Monte Carlo approach (HMC) also known as hybrid Monte Carlo~\citep{duane_1987, neal_mcmc_2012}.

To avoid evaporation events, where atoms detach from the cluster a confining potential was added to the systems. 
We used the same soft potential as~\citet{nigra_combining_2005} which is given as follows.
\begin{equation}
	V(\mathbf{r}) = \sum_{i=1}^N \epsilon \left( \frac{\Vert \vec{r}_i - \vec{r}_{cm}\Vert}{r_c} \right)^{20}
\end{equation}
With $\vec{r}_{cm}$ being the center of mass and $r_c$ the radius of the confining potential.
We experimentally found $r_c=3.5\sigma$ to be a good choice for $LJ_{38}$ and $r_c=4\sigma$ for $LJ_{75}$. 
A good choice of $r_c$ has to prevent atoms from escaping without influencing the energy of the cluster too much. 
A soft potential was used because the derivatives/forces were needed for the Hamiltonian dynamics.

We compared the results of our method to parallel tempering~\citep{swendsen_replica_1986, neirotti_phase_2000, earl_parallel_2005}.
We used a geometric distribution of the temperatures as proposed by~\citet{kofke_acceptance_2002}.
In our simulation swaps were attempted every 10 Monte Carlo steps between adjacent temperatures. They were accepted with a rate of 16-19 percent for $LJ_{38}$ and between 13 and 17 percent for $LJ_{75}$ which is close to the optimal acceptance rate of 20\% proposed by~\citet{rathore_optimal_2005}. 
Swaps were performed in an alternating manner between pairs of subsequent temperatures e.g. after the first ten steps swaps were attempted between pairs 1-2, 3-4, 5-6, ... and then after ten more steps pairs 2-3, 4-5, 6-7, etc. were used.

The heat capacity was calculated using the following equation.
\begin{equation}
    C_V(T) = \frac{3}{2} + \frac{1}{N T^2} \left( {\langle E^2 \rangle}_T - {\langle E \rangle}_T^2 \right)
\end{equation}
With ${\langle \cdot \rangle}_T$ representing expectation values over the Boltzmann distribution at temperature $T$.

To obtain smooth plots of the heat capacity we used the re-weighting scheme proposed by~\citet{sharapov_solidsolid_2007} to interpolate between the different temperatures.

\subsection{Lennard-Jones 38}
The most studied Lennard-Jones cluster is presumably the one consisting of 38 atoms ($LJ_{38}$) which is known for its two funnel energy landscape that almost completely prevents ergodic sampling using conventional Monte Carlo methods.

One funnel ends in the global minimum which is a face-centered-cubic truncated octahedral structure.
The other funnel ends in the second lowest minimum, which is an incomplete Mackay icosahedron.
These two funnels are separated by a high energy barrier with a transition state energy of $4.219\epsilon$ above the ground state energy~\citep{doye_double-funnel_1999} which is almost impossible to overcome at low temperatures.

\begin{figure}
\centering
\includegraphics[width=\linewidth]{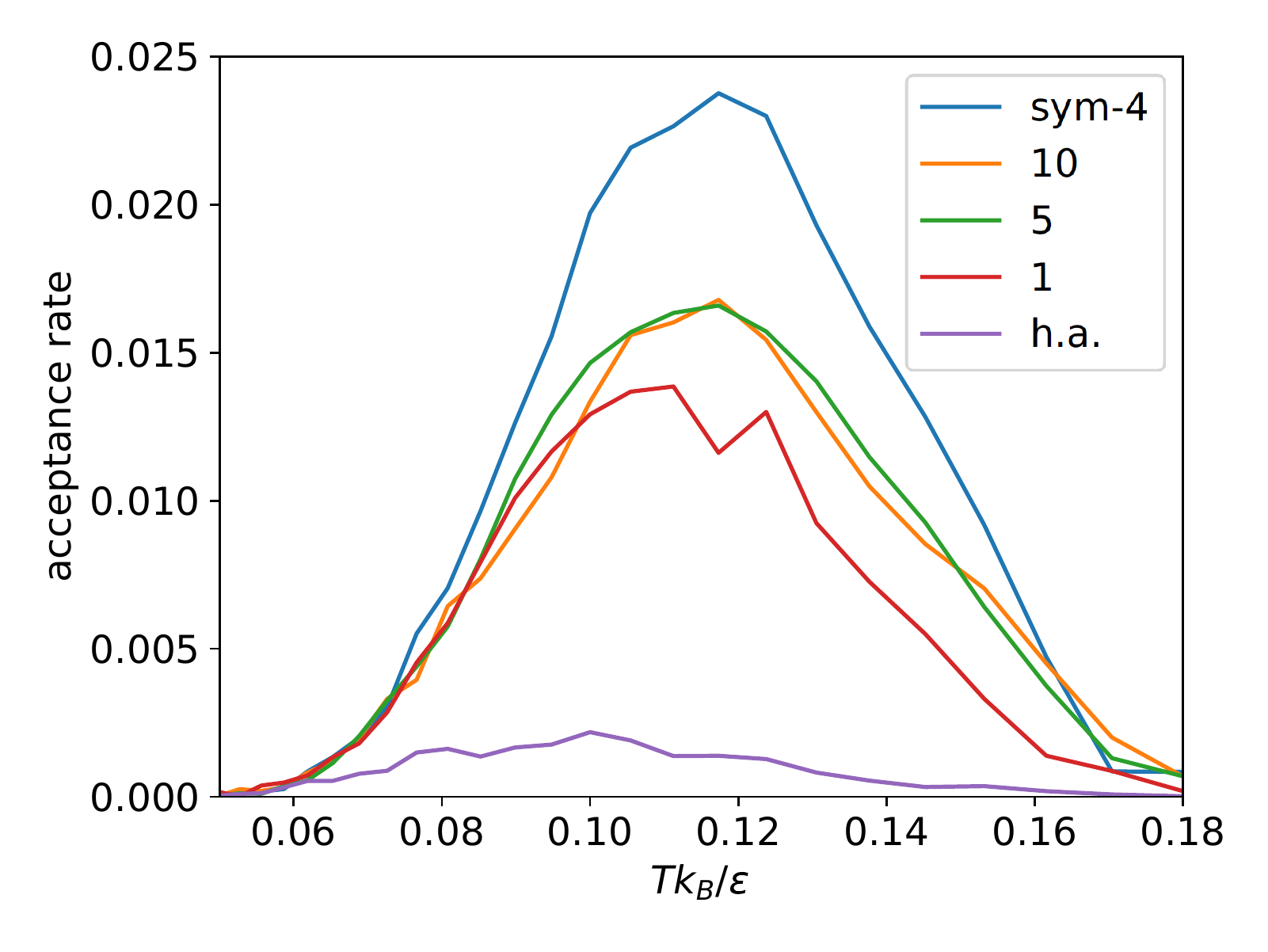}
\caption{Acceptance rate of funnel hopping moves in $LJ_{38}$ plotted against temperature. The numbers represent the number of Gaussians used in the Gaussian Mixtures. For the line labelled h.a. the harmonic approximation was used and for the line labelled sym-4 an extended version of the EM algorithm was used to fit a symmetric Gaussian Mixture with 4 Gaussias per symmetry.}
\label{fig:lj38-acc}
\end{figure}

Gaussian mixtures were fit for the ten lowest local minima (stereoisomers were counted as one) using $2 \times 10^5$ samples.
The acceptance rates achieved are shown in Fig.~\ref{fig:lj38-acc}.

As one can see in Fig.~\ref{fig:lj38-acc} going from 5 to 10 Gaussians did not increase the acceptance rate.
We suspect that this is due to the number of samples not being sufficient for the high numbers of parameters that have to be fitted. In this case we have 5995 free parameters per Gaussian. $n (n+1)/2$ from the covariance matrix, \cc{$n$} from the mean and one $a_i^k$ with $n$ being the dimensionality of the Gaussian which is $3 \cdot 38 - 6$ in this case.

To further increase the quality of the fit without having to generate more samples we developed a method to incorporate the high symmetry of the local minima into the fitting procedure. 
In appendix~\ref{sec:sym-em} an outline of the algorithm can be found. 
The acceptance rate for the Monte Carlo run using a symmetric Gaussian Mixture is labeled \textit{sym-4} in Fig.~\ref{fig:lj38-acc}. The Gaussian Mixture for this fit consists of 4 Gaussians per symmetry of the local minima.

\begin{figure}
\centering
\includegraphics[width=\linewidth]{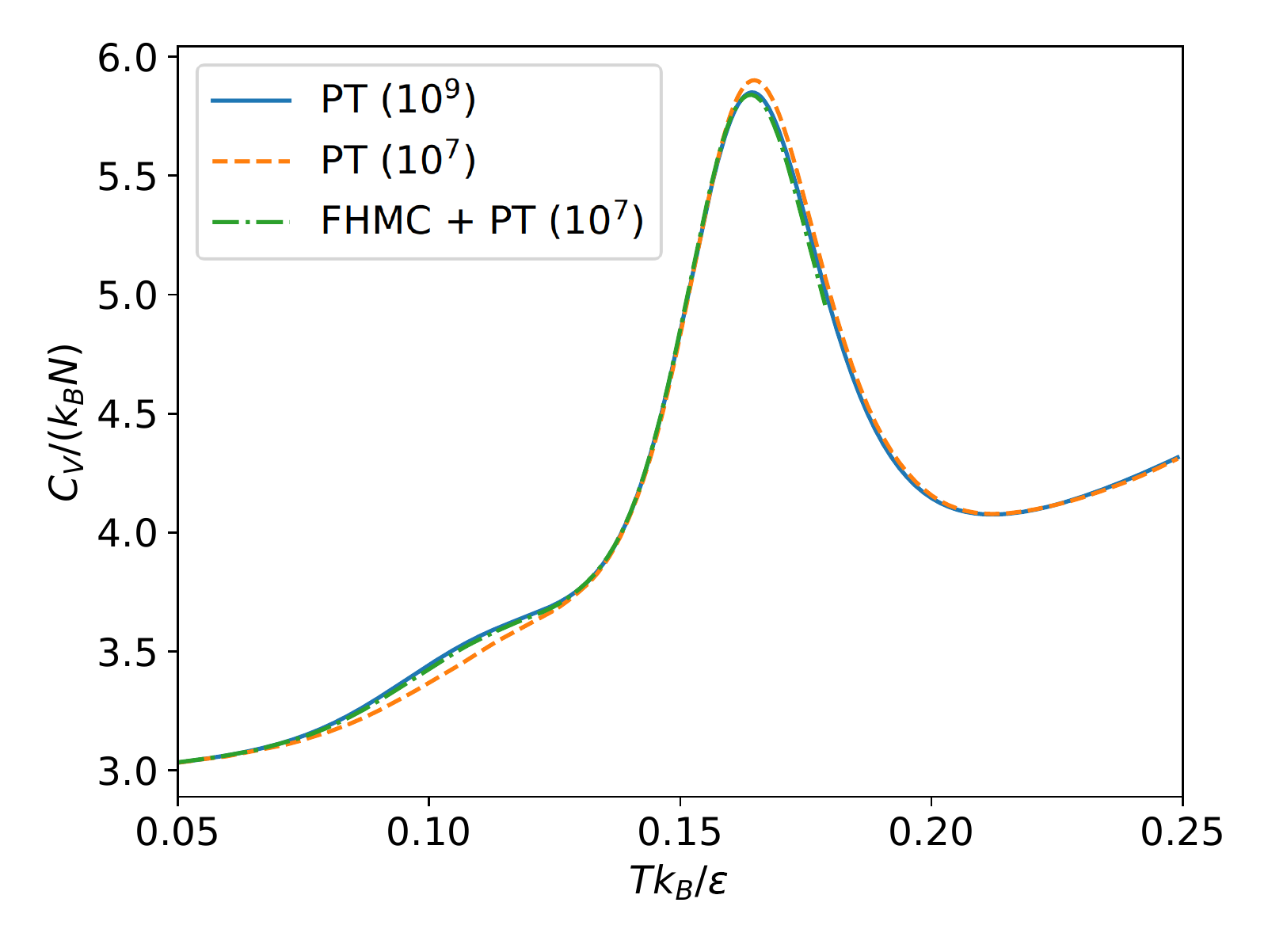}
\caption{Heat capacity of $LJ_{38}$ calculated with our method compared to the result obtained using parallel tempering after different numbers of steps. }
\label{fig:lj38-cv} 
\end{figure}

This fit was then used to calculate the heat capacity of $LJ_{38}$ using \methodnameshort~in combination with parallel tempering. The result is shown in Fig.~\ref{fig:lj38-cv} where it is compared to a reference calculation using parallel tempering with $10^9$ steps. 
In our implementation of both the simple parallel tempering and the \methodnameshort~method one step of the Hamiltonian Monte Carlo algorithm required 25 energy and force evaluations.
The curve using both methods in combination was obtained after $10^7$ steps. While this result is in agreement with the reference the result obtained with parallel tempering alone using the same number of steps is clearly not converged.

\begin{figure}
\centering
\includegraphics[width=\linewidth]{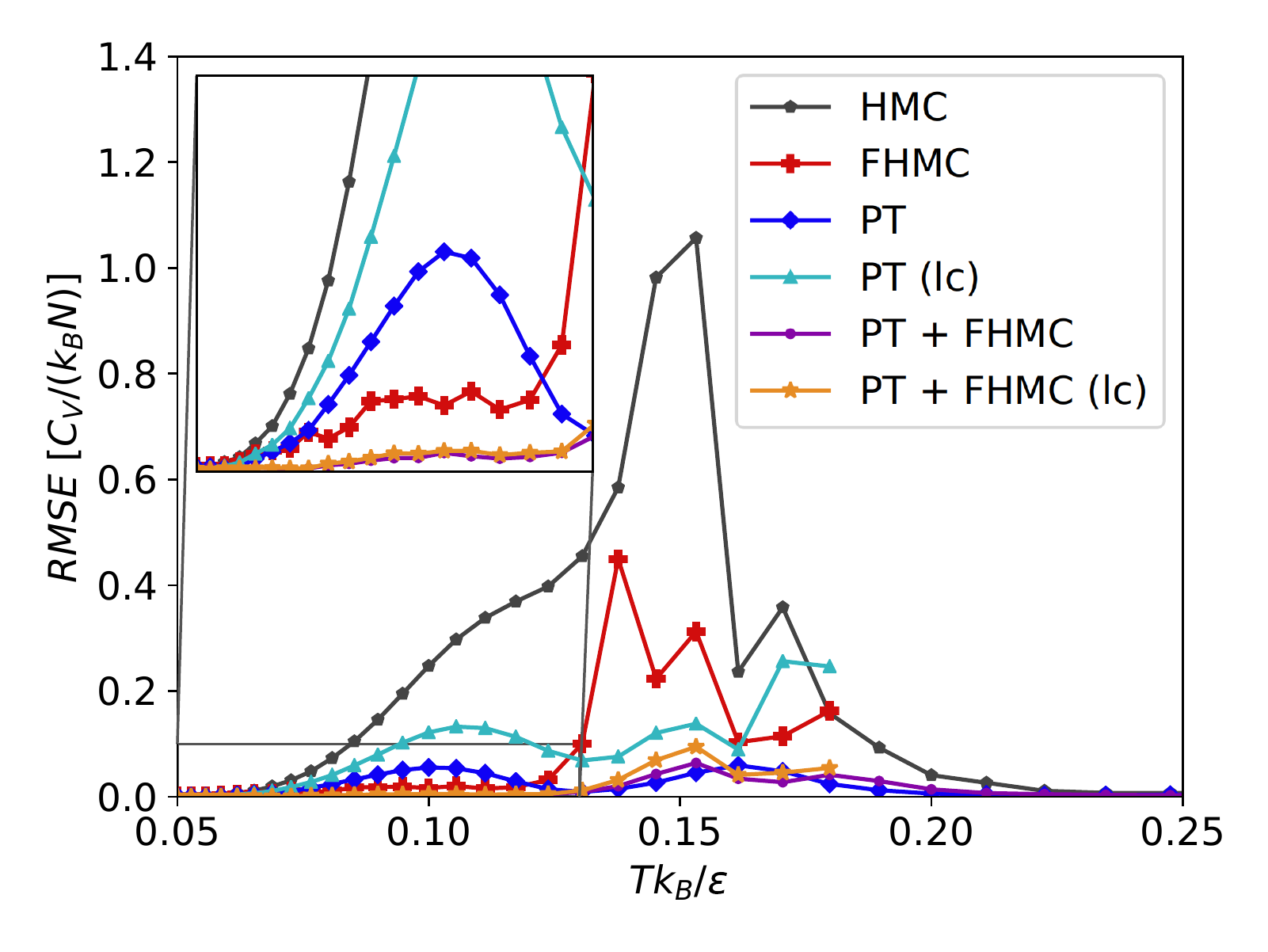}
\caption{Root mean squared error of the heat capacity of $LJ_{38}$ calculated with Hamiltonian Monte Carlo (HMC), parallel tempering (PT), \methodname~(\methodnameshort), PT and \methodnameshort~in combination once with a regular and once with a lower maximum temperature (lc).}
\label{fig:lj38-err-subplot}
\end{figure}

To assess the convergence properties of our method we repeated the calculation of the heat capacity ten times with both methods individually and combined using $10^7$ steps. We then calculated the root mean squared error (RMSE) with respect to the reference obtained with $10^9$ parallel tempering steps. The resulting RMSEs are shown in Fig.~\ref{fig:lj38-err-subplot}.

The results show that our method alone, outperforms parallel tempering at the lower temperature range up to $T=0.11\epsilon/k_B$.
In this range the number of accessible minima is low enough so that they are well covered by the darting sites. 
In this special case \Methodname~can be used to perform ergodic sampling using only one simulation at a single temperature, reducing the computational effort by several orders of magnitude compared to parallel tempering simulations where a whole range of temperatures has to be simulated. 
\cc{We found that if we included only the lowest minimum in each funnel the \methodnameshort~calculation did not converge within $10^7$ steps. If combined with parallel tempering however the convergence was only slightly slower than with 10 minima.}

At higher temperatures additional minima become relevant that are not included into the \methodnameshort~scheme and have to be reached by standard Monte Carlo moves which can slow down convergence~\citep{neirotti_phase_2000}.

One major drawback of parallel tempering is that a large range of temperatures has to be simulated with the maximum temperature being high enough so that the highest energy barriers can be crossed by the Monte Carlo simulation. In our experiments we chose a maximum temperature of $0.4\epsilon/k_B$ for parallel tempering while the maximum temperature for the \methodnameshort~simulations can be chosen arbitrarily because each simulation is performed independently of the others. 
For our \methodnameshort~simulations we chose a maximum temperature of $0.18\epsilon/k_B$. 
Parallel tempering simulations with this maximum temperature did not converge to the correct result. When we combined parallel tempering with our method however convergence could be achieved. In this case the Funnel Hopping moves allow the simulation to cross the highest barriers while parallel tempering enables the crossing of the lower barriers between basins within a funnel that are not included into the \methodnameshort~scheme. 
Using both methods in combination allows therfore to use a significantly lower cutoff temperature than with standalone parallel tempering. It combines the best of both methods by using parallel tempering to skip barriers inside funnels and \methodname~to move between different funnels, leading to improved sampling capabilities across the whole temperature range which can also be seen in Fig.~\ref{fig:lj38-err-subplot}.

\subsection{Lennard-Jones 75}

As a final test we applied \methodname~to an even more challenging system. namely the 75 atom Lennard-Jones cluster ($LJ_{75}$). Similar to the 38 atom cluster, its energy landscape also consists of two major funnels, one ending in the global optimum which has a decahedral structure and the other one ending in the second lowest local minimum which has an icosahedral structure. These two minima are separated by a barrier that lies $8.69 \epsilon$ above the ground state energy. This barrier is over $3\epsilon$ higher than any other barrier between the 250 lowest minima~\citep{doye_evolution_1999}.
Unlike in the case of $LJ_{38}$ the peak in the heat capacity caused by the solid-solid transition is well separated from the melting peak.

The very high barrier between the two funnels makes ergodic sampling of this system particularly difficult. It seems that parallel tempering alone is not enough to obtain converged results for $LJ_{75}$~\citep{sharapov_solidsolid_2007, nigra_combining_2005}. Our own calculation using parallel tempering with Hamiltonian Monte Carlo did not converge after $5\times10^8$ steps per temperature ($5\times10^{11}$ energy and force evaluations in total).
Because of the high energy barrier between the two funnels transitions are limited to the very high temperature range of the parallel tempering simulation. At these temperatures the accessible configuration space is extremely large causing the transition between the funnels to be particularly rare.

By combining parallel tempering with \methodname~transitions between the two funnels become possible already at low temperature.

We used Funnel Hopping Monte Carlo in combination with parallel tempering to calculate the heat capacity of $LJ_{75}$.
The two lowest minima were included into the FHMC scheme to facilitate the crossing of the high inter funnel barrier.
We used our version of the EM algorithm to fit Gaussian mixtures of three Gaussians per symmetry using \cc{$2 \times 10^5$} samples per local minimum. 

\begin{figure}
\centering
\includegraphics[width=\linewidth]{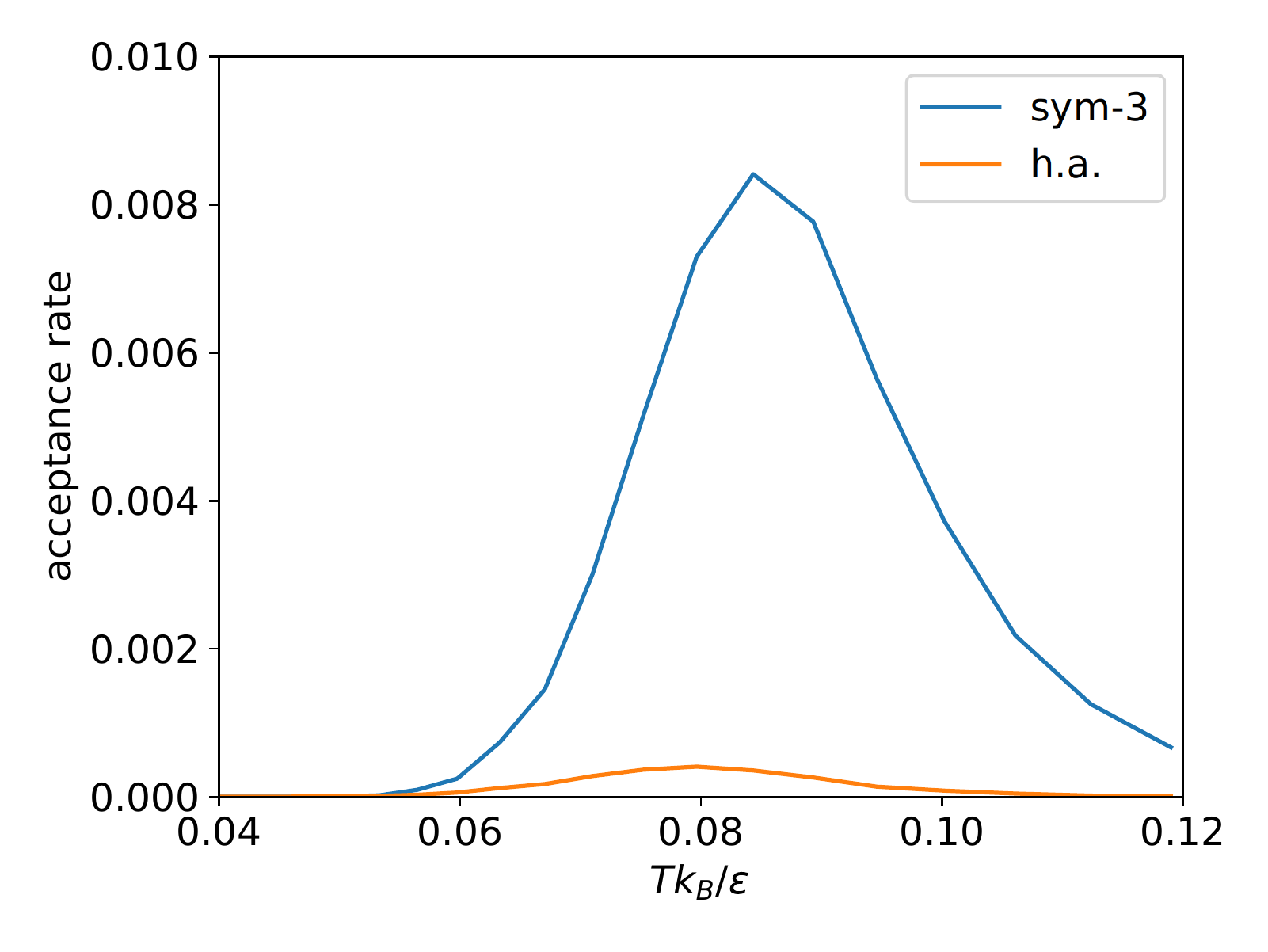}
\caption{Acceptance rate of funnel hopping moves in $LJ_{75}$ plotted against temperature. The result obtained using the Gaussian mixtures consisting of 3 Gaussians per symmetry, are labeled with \textit{sym-3} while the result obtained using the harmonic approximation are labeled \textit{h.a.}.}
\label{fig:lj75-acc}
\end{figure}

FHMC moves were included with a probability of $0.1$ up to a temperature of $0.119\epsilon/k_B$ above which the acceptance rate of the moves decays to almost zero. 
\cc{A total of 40 parallel tempering replicas were used, which were run in parallel each on one CPU core. The lower 20 of the replicas included \methodnameshort~moves. Because we included two local minima into the \methodnameshort~scheme, 40 Gaussian mixtures were fit.}
The resulting acceptance rates are shown in figure~\ref{fig:lj75-acc}.
The fitted Gaussian mixtures outperform the harmonic approximation in terms of the acceptance rate of the proposed moves by about a factor of 20.

Parallel tempering swaps were again included after every 10 steps. Samples were collected after an equilibration period of $10^5$ steps.

\begin{figure}
\centering
\includegraphics[width=\linewidth]{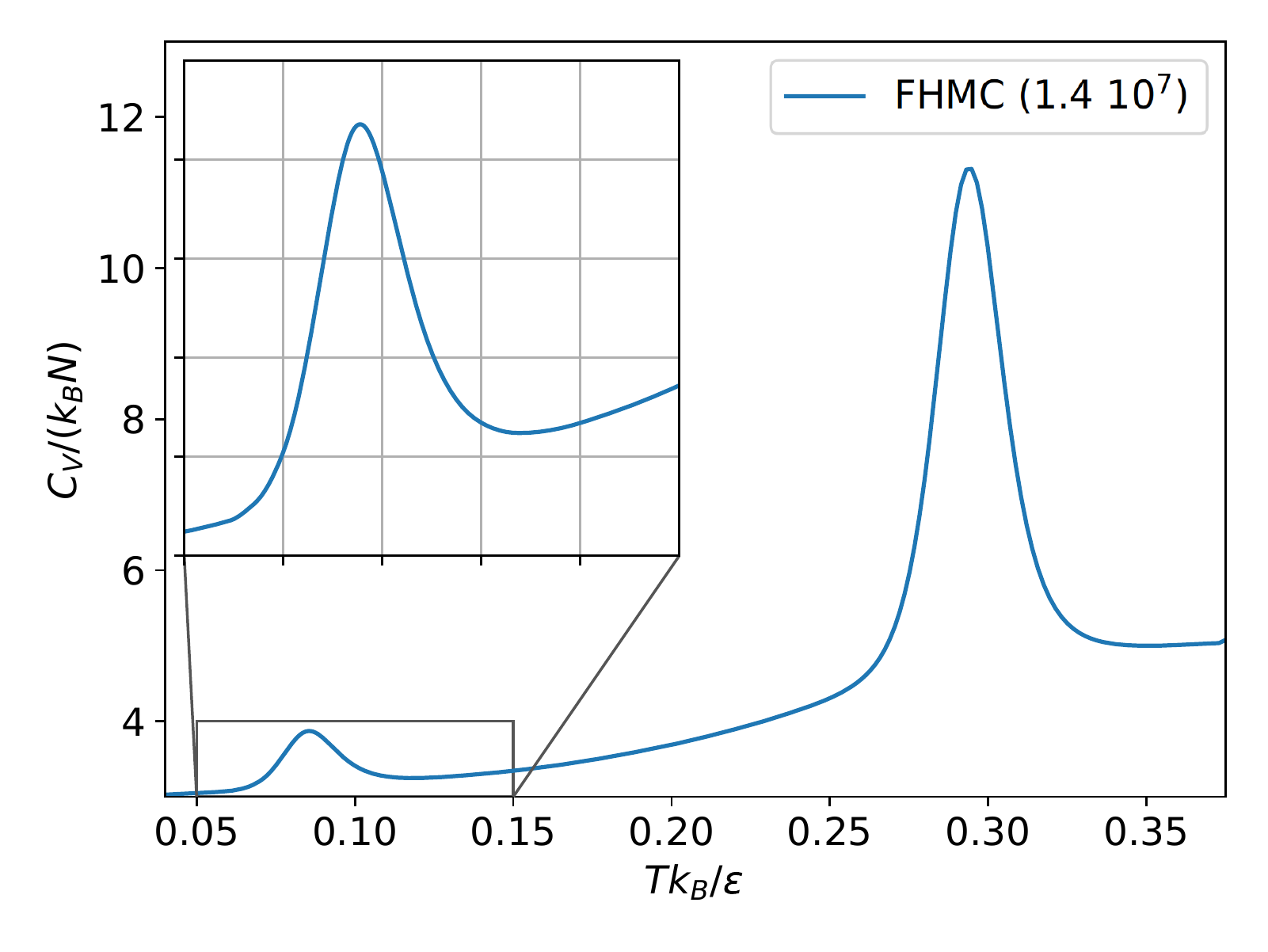}
\caption{Heat capacity of $LJ_{75}$ calculated with~\Methodname.}
\label{fig:lj75-fhmc}
\end{figure}

The obtained heat capacity after $1.4 \times 10^7$ steps is shown in Fig.~\ref{fig:lj75-fhmc}.
We identified the peak of the heat capacity corresponding to the solid-solid transition at a temperature of $0.085\epsilon/k_B$.
This is slightly higher than the result of $0.083\epsilon/k_B$ reported by~\citet{sharapov_solidsolid_2007}.
To explain this minor discrepancy we ran several simulation, initialized with the second lowest instead of the lowest minimum, with a larger confining radius as well as with a longer equilibration period. However the results of all these calculations gave the same numerical value for the peak.

\begin{figure}
\centering
\includegraphics[width=\linewidth]{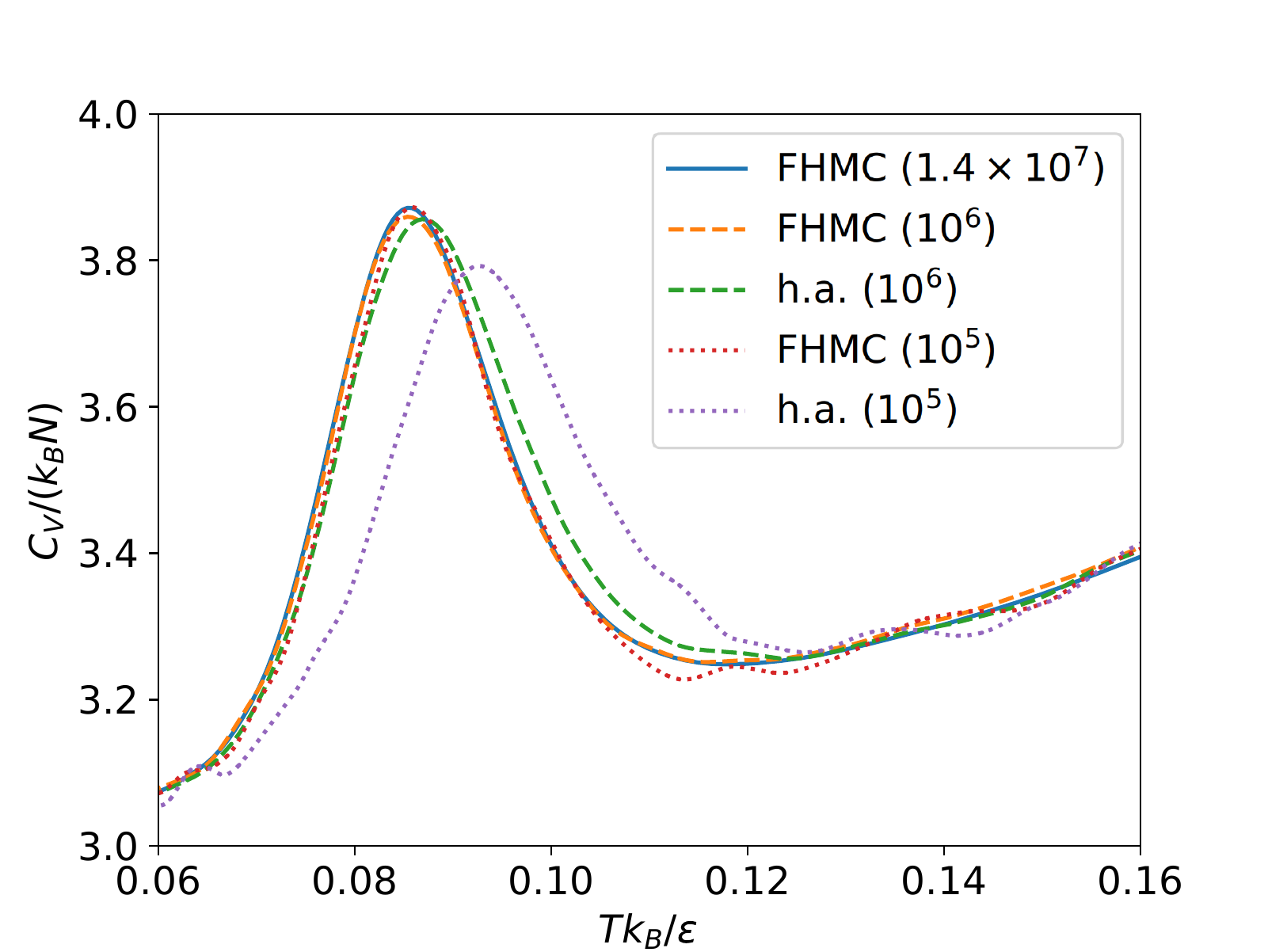}
\caption{Low temperature heat capacity peak of $LJ_{75}$ calculated with our method using a Gaussian mixture and the harmonic approximation. The number given in the legend indicates the number of HMC steps during which samples were collected (1 step = 25 energy and force evaluations per temperature).}
\label{fig:lj75-conv}
\end{figure}

In Fig.~\ref{fig:lj75-conv} the peak corresponding to the low temperature solid-solid transition is shown again and compared to the results obtained with our method and with a run where the harmonic approximation was used instead of fitted Gaussians, both after $10^5$ and $10^6$ steps.
After $10^6$ steps the \methodnameshort~calculation is converged to the final result after $1.4\times10^7$ steps up to a very high precision while the result from the harmonic approximation is still significantly shifted towards the right. Even after only $10^5$ steps the \methodnameshort~ calculation is already converged to a result where the heat capacity peak is in good qualitative agreement with the converged result.
These results clearly indicate that the rate at which the simulation jumps between the two funnels is the limiting factor for the convergence of the Monte Carlo simulation.

Hence, using \Methodname~we were able to obtain a converged result after only $10^6$ steps ($3.25\times10^7$ energy and force evaluations per temperature or $1.3\times10^9$ in total, including sample generation for the Gaussian mixtures as well as the equilibration part).
This is almost 100 times less than the $3\times10^9$ energy evaluations per temperature reported by~\citet{sharapov_low-temperature_2007} where an auxiliary harmonic superposition systems was used and more than 100 times less then the $4\times10^{11}$ energy evaluations in total reported by~\citet{martiniani_superposition_2014} where the approximate SENS method was employed.
\cc{The basin-sampling method~\citep{wales_surveying_2013} used $0.27\times10^9$ energy evaluations per replica resulting in a total of $8.64\times10^9$ energy evaluations for all 32 replicas combined  not including the energy evaluations required for the minimizations in the final BS phase. This  is significantly more than in the \methodnameshort~method.}

\section{conclusion}
With \methodname~we developed a new tool to overcome broken ergodicity by exploiting precomputed knowledge about the energy landscape into the Monte Carlo simulation.
Our method generates an accurate approximation to the Boltzmann distribution even for anharmonic systems.
This allows us to propose good moves between different funnels that have a high chance of being accepted by the Monte Carlo algorithm. Using Gaussian mixtures allows for a systematic improvement of the proposed moves by increasing the number of samples used for fitting and the number of Gaussians in the Gaussian mixture. With our newly developed variant of the expectation-maximization algorithm we are able to take advantage of the high symmetry present in many local minima which results in an even better fit of the Gaussian mixtures.
With our fits we were able to achieve acceptance rates about twenty times higher than with the harmonic approximation. 
We observed that the convergence of the Monte Carlo simulation is limited by the rate at which the simulation is able to transition between the different funnels and therefore directly dependent on the acceptance rate of the inter funnel moves.

If the temperature of interest is low enough so that only a limited number of basins are accessible and if it is possible to include all of them into the algorithm, \methodname~can be performed at a single temperature whereas parallel tempering requires many auxiliary simulations at higher temperatures.

We also showed that by combining our method within the parallel tempering scheme the maximum simulation temperature, can be significantly reduced which allows to avoid unnecessary calculations resulting in a reduced computational cost. Also the convergence of the simulation is sped up massively as the \methodname~moves help the simulation to cross the highest barriers between different funnels very efficiently.

Using \methodname~we were able to obtain the heat capacity of the 75 atom Lennard-Jones, a notoriously difficult system, known to suffer from a particularly strong broken ergodicity. Nevertheless, using only $1.3\times10^9$ energy and force evaluations in total we could obtain converged results. This number 
of evaluations is \cc{significantly} less than the number required by existing state of the art methods.
We also observed that the results were already in good qualitative agreement after only about $10^8$ energy and force evaluations. 

\begin{acknowledgments}
This work was partly supported by the NCCR MARVEL and funded by the SNF. 
J.A.F. acknowledges support from the SNF within a joint DFG/SNF project. 
Computational resources were provided by the Swiss National Supercomputing Centre (CSCS) in Lugano under the project s963 are gratefully acknowledged.  Calculations were also performed at sciCORE (\href{http://scicore.unibas.ch/}{http://scicore.unibas.ch/}) scientific computing center at University of Basel.

\end{acknowledgments}

\appendix
\section{Fitting symmetric Gaussian mixtures}\label{sec:sym-em}

Low energy configurations of clusters often exhibit a high degree of symmetry.
This is especially the case for the Lennard-Jones 38 cluster where the ground state has 24 rotational symmetries as well as an inversion symmetry resulting in a total of 48 symmetries.

These symmetries will also be present in the Boltzmann distribution which we approximate using the Gaussian mixtures. 
By constraining the Gaussian mixtures to have the same symmetries as the local minima, the number of free parameters can be reduced, which results in an increased quality of fit with the same number of training samples used.
We therefore developed the following variant of the expectation-maximization algorithm.

In a first step we determine all rotation and inversion symmetries of the configuration. 
For that the configuration is first rotated randomly, then the alignment algorithm described in section~\ref{sec:minRmsd} is used to align the rotated structure to the original configuration. All distinct assignments with an RMSD of zero correspond to a symmetry operation. The procedure is then repeated with the structure being inverted so that all symmetries that include an inversion can be found. Alternatively the symmetries can also be detected using a more efficient code, such as for example libmsym~\citep{johansson_automatic_2017}.

We then define a symmetric Gaussian mixture by replicating a normal Gaussian mixture for each of the symmetry operations. 
\begin{equation}\label{eq:symgmm}
    	q^{sym}(\mathbf{r}) = \sum_{j=1}^{N_{sym}} \sum_{k=1}^m a^k \mathcal{N}_j^k(\mathbf{r})
\end{equation}
Here $N_{sym}$ is the number of symmetries, $m$ is the number of Gaussians per symmetry and $\mathcal{N}_j^k$ is the $k$th Gaussian under the $j$th symmetry transformation. Similar to the non symmetric Gaussian mixture the $a_k$s are weights for the individual Gaussians. Because each Gaussian is replicated $N_{sym}$ times the $a_k$ have to sum up to $1/N_{sym}$.

Each symmetry operation consists of a rotation represented by a rotation matrix $\mathrm{R}$, a permutation represented by a permutation matrix $\mathrm{P}$, and optionally an inversion. To apply a symmetry transformation to the $3N-6$ dimensional vectors we first have to transform them back to the $3N$ dimensional space. Then the rotation, permutation and inversion are applied before transforming back to the $3N-6$ dimensional coordinates. 
Combining all of these operations yields the matrix $\mathrm{M}$
\begin{equation}
    \mathrm{M} = \mathrm{B}^\top \mathrm{P} \mathrm{Q} \mathrm{I} \mathrm{B}
\end{equation}
With $\mathrm{B}$ being a $3N\times3N-6$ matrix with its columns consisting of the $3N-6$ basis vectors $\mathbf{B}^i$, $\mathrm{Q}$ being a block diagonal matrix with the rotation matrix $\mathrm{R}$ repeated $N$ times along its diagonal and $\mathrm{I}$ being the identity matrix $\mathbbm{1}$, or $-\mathbbm{1}$ if an inversion is applied. The Gaussian $\mathcal{N}_j^k$ is hence defined by the mean $\mu_j^k = \mathrm{M}_j \mu^k$ and the covariance $\Sigma_j^k = \mathrm{M}_j\Sigma^k\mathrm{M}j^\top$.
The symmetric Gaussian mixture model is therefore parametrized by $m$ weights $a^k$, $m$ mean vectors, and $m$ covariance matrices.

To fit this symmetric Gaussian mixture we modified the original expectation-maximization algorithm in the following way.
During the expectation part of each iteration we first construct the full symmetric Gaussian mixture as it is given by equation~\ref{eq:symgmm}. We then calculate the weights $y_i^{jk}$ for each sample $x_i$ in the same way as in the original algorithm.

\begin{equation}
    y_i^{jk} = \frac{a^k \mathcal{N}_j^k(x_i)}
{\sum_{j=1}^{N_{sym}} \sum_{k=1}^{m} a^k \mathcal{N}_j^k(x_i)}
\end{equation}
For the parameter estimation in the maximization step of the algorithm we apply the inverse symmetry transformations $\mathrm{M}^\top$ to the samples. The weight calculated for sample $i$ with Gaussian $k$ transformed with symmetry $j$ is now used on the sample transformed with $\mathrm{M}_j^\top$ to estimate the parameters $\mu^k$ and $\Sigma^k$

\begin{samepage}
\begin{equation}
    \mu^k = \frac{\sum_{j=1}^{N_{sym}}\sum_{i=1}^{N} y_i^{jk} \mathrm{M}_j^\top x_i}
                 {\sum_{ij} y_i^{jk}} 
\end{equation}
\begin{equation}
    \Sigma^k = \frac{1}{\sum_{ij} y_i^{jk}} 
            \sum_{j=1}^{N_{sym}}\sum_{i=1}^{N} y_i^{jk} (\mathrm{M}_j^\top x_i - \mu^k)(\mathrm{M}_j^\top x_i - \mu^k)^\top
\end{equation}
\end{samepage}

As in the original version of the algorithm the expectation and maximization steps are repeated alternately until convergence is achieved.

With this modified version of the expectation-maximization algorithm we were able to achieve significantly better fits and hence in higher performance of our~\methodname~algorithm whenever symmetries were present in any of the local minima used.

\bibliography{fhmc}

\begin{thebibliography}{49}%
\makeatletter
\providecommand \@ifxundefined [1]{%
 \@ifx{#1\undefined}
}%
\providecommand \@ifnum [1]{%
 \ifnum #1\expandafter \@firstoftwo
 \else \expandafter \@secondoftwo
 \fi
}%
\providecommand \@ifx [1]{%
 \ifx #1\expandafter \@firstoftwo
 \else \expandafter \@secondoftwo
 \fi
}%
\providecommand \natexlab [1]{#1}%
\providecommand \enquote  [1]{``#1''}%
\providecommand \bibnamefont  [1]{#1}%
\providecommand \bibfnamefont [1]{#1}%
\providecommand \citenamefont [1]{#1}%
\providecommand \href@noop [0]{\@secondoftwo}%
\providecommand \href [0]{\begingroup \@sanitize@url \@href}%
\providecommand \@href[1]{\@@startlink{#1}\@@href}%
\providecommand \@@href[1]{\endgroup#1\@@endlink}%
\providecommand \@sanitize@url [0]{\catcode `\\12\catcode `\$12\catcode
  `\&12\catcode `\#12\catcode `\^12\catcode `\_12\catcode `\%12\relax}%
\providecommand \@@startlink[1]{}%
\providecommand \@@endlink[0]{}%
\providecommand \url  [0]{\begingroup\@sanitize@url \@url }%
\providecommand \@url [1]{\endgroup\@href {#1}{\urlprefix }}%
\providecommand \urlprefix  [0]{URL }%
\providecommand \Eprint [0]{\href }%
\providecommand \doibase [0]{http://dx.doi.org/}%
\providecommand \selectlanguage [0]{\@gobble}%
\providecommand \bibinfo  [0]{\@secondoftwo}%
\providecommand \bibfield  [0]{\@secondoftwo}%
\providecommand \translation [1]{[#1]}%
\providecommand \BibitemOpen [0]{}%
\providecommand \bibitemStop [0]{}%
\providecommand \bibitemNoStop [0]{.\EOS\space}%
\providecommand \EOS [0]{\spacefactor3000\relax}%
\providecommand \BibitemShut  [1]{\csname bibitem#1\endcsname}%
\let\auto@bib@innerbib\@empty
\bibitem [{\citenamefont {Stillinger}\ and\ \citenamefont
  {Weber}(1982)}]{stillinger_1982}%
  \BibitemOpen
  \bibfield  {author} {\bibinfo {author} {\bibfnamefont {F.~H.}\ \bibnamefont
  {Stillinger}}\ and\ \bibinfo {author} {\bibfnamefont {T.~A.}\ \bibnamefont
  {Weber}},\ }\bibfield  {title} {\enquote {\bibinfo {title} {Hidden structure
  in liquids},}\ }\href {\doibase 10.1103/PhysRevA.25.978} {\bibfield
  {journal} {\bibinfo  {journal} {Phys. Rev. A}\ }\textbf {\bibinfo {volume}
  {25}},\ \bibinfo {pages} {978--989} (\bibinfo {year} {1982})}\BibitemShut
  {NoStop}%
\bibitem [{\citenamefont {Calvo}, \citenamefont {Doye},\ and\ \citenamefont
  {Wales}(2002)}]{calvo_equilibrium_2002}%
  \BibitemOpen
  \bibfield  {author} {\bibinfo {author} {\bibfnamefont {F.}~\bibnamefont
  {Calvo}}, \bibinfo {author} {\bibfnamefont {J.}~\bibnamefont {Doye}}, \ and\
  \bibinfo {author} {\bibfnamefont {D.}~\bibnamefont {Wales}},\ }\bibfield
  {title} {{\selectlanguage {english}\enquote {\bibinfo {title} {Equilibrium
  properties of clusters in the harmonic superposition approximation},}\
  }}\href {\doibase 10.1016/S0009-2614(02)01550-6} {\bibfield  {journal}
  {\bibinfo  {journal} {Chemical Physics Letters}\ }\textbf {\bibinfo {volume}
  {366}},\ \bibinfo {pages} {176--183} (\bibinfo {year} {2002})}\BibitemShut
  {NoStop}%
\bibitem [{\citenamefont {Hastings}(1970)}]{hastings_monte_1970}%
  \BibitemOpen
  \bibfield  {author} {\bibinfo {author} {\bibfnamefont {W.~K.}\ \bibnamefont
  {Hastings}},\ }\bibfield  {title} {{\selectlanguage {english}\enquote
  {\bibinfo {title} {Monte {Carlo} sampling methods using {Markov} chains and
  their applications},}\ }}\href {\doibase 10.1093/biomet/57.1.97} {\bibfield
  {journal} {\bibinfo  {journal} {Biometrika}\ }\textbf {\bibinfo {volume}
  {57}},\ \bibinfo {pages} {97--109} (\bibinfo {year} {1970})}\BibitemShut
  {NoStop}%
\bibitem [{\citenamefont {Wales}(2012)}]{wales_decoding_2012}%
  \BibitemOpen
  \bibfield  {author} {\bibinfo {author} {\bibfnamefont {D.~J.}\ \bibnamefont
  {Wales}},\ }\bibfield  {title} {{\selectlanguage {english}\enquote {\bibinfo
  {title} {Decoding the energy landscape: extracting structure, dynamics and
  thermodynamics},}\ }}\href {\doibase 10.1098/rsta.2011.0208} {\bibfield
  {journal} {\bibinfo  {journal} {Philosophical Transactions of the Royal
  Society A: Mathematical, Physical and Engineering Sciences}\ }\textbf
  {\bibinfo {volume} {370}},\ \bibinfo {pages} {2877--2899} (\bibinfo {year}
  {2012})}\BibitemShut {NoStop}%
\bibitem [{\citenamefont {Wales}(2010)}]{wales_energy_2010}%
  \BibitemOpen
  \bibfield  {author} {\bibinfo {author} {\bibfnamefont {D.~J.}\ \bibnamefont
  {Wales}},\ }\bibfield  {title} {{\selectlanguage {english}\enquote {\bibinfo
  {title} {Energy landscapes: some new horizons},}\ }}\href {\doibase
  10.1016/j.sbi.2009.12.011} {\bibfield  {journal} {\bibinfo  {journal}
  {Current Opinion in Structural Biology}\ }\textbf {\bibinfo {volume} {20}},\
  \bibinfo {pages} {3--10} (\bibinfo {year} {2010})}\BibitemShut {NoStop}%
\bibitem [{\citenamefont {Torrie}\ and\ \citenamefont
  {Valleau}(1977)}]{umbrella-sampling}%
  \BibitemOpen
  \bibfield  {author} {\bibinfo {author} {\bibfnamefont {G.~M.}\ \bibnamefont
  {Torrie}}\ and\ \bibinfo {author} {\bibfnamefont {J.~P.}\ \bibnamefont
  {Valleau}},\ }\bibfield  {title} {\enquote {\bibinfo {title} {Nonphysical
  sampling distributions in monte carlo free-energy estimation: Umbrella
  sampling},}\ }\href@noop {} {\bibfield  {journal} {\bibinfo  {journal}
  {Journal of Computational Physics}\ }\textbf {\bibinfo {volume} {23}},\
  \bibinfo {pages} {187--199} (\bibinfo {year} {1977})}\BibitemShut {NoStop}%
\bibitem [{\citenamefont {Laio}\ and\ \citenamefont
  {Parrinello}(2002)}]{laio2002escaping}%
  \BibitemOpen
  \bibfield  {author} {\bibinfo {author} {\bibfnamefont {A.}~\bibnamefont
  {Laio}}\ and\ \bibinfo {author} {\bibfnamefont {M.}~\bibnamefont
  {Parrinello}},\ }\bibfield  {title} {\enquote {\bibinfo {title} {Escaping
  free-energy minima},}\ }\href@noop {} {\bibfield  {journal} {\bibinfo
  {journal} {Proceedings of the National Academy of Sciences}\ }\textbf
  {\bibinfo {volume} {99}},\ \bibinfo {pages} {12562--12566} (\bibinfo {year}
  {2002})}\BibitemShut {NoStop}%
\bibitem [{\citenamefont {Berg}\ and\ \citenamefont
  {Neuhaus}(1991)}]{multicanonical-algo}%
  \BibitemOpen
  \bibfield  {author} {\bibinfo {author} {\bibfnamefont {B.~A.}\ \bibnamefont
  {Berg}}\ and\ \bibinfo {author} {\bibfnamefont {T.}~\bibnamefont {Neuhaus}},\
  }\bibfield  {title} {\enquote {\bibinfo {title} {Multicanonical algorithms
  for first order phase transitions},}\ }\href@noop {} {\bibfield  {journal}
  {\bibinfo  {journal} {Physics Letters B}\ }\textbf {\bibinfo {volume}
  {267}},\ \bibinfo {pages} {249--253} (\bibinfo {year} {1991})}\BibitemShut
  {NoStop}%
\bibitem [{\citenamefont {Wang}\ and\ \citenamefont
  {Landau}(2001)}]{wang_efficient_2001}%
  \BibitemOpen
  \bibfield  {author} {\bibinfo {author} {\bibfnamefont {F.}~\bibnamefont
  {Wang}}\ and\ \bibinfo {author} {\bibfnamefont {D.~P.}\ \bibnamefont
  {Landau}},\ }\bibfield  {title} {{\selectlanguage {english}\enquote {\bibinfo
  {title} {Efficient, {Multiple}-{Range} {Random} {Walk} {Algorithm} to
  {Calculate} the {Density} of {States}},}\ }}\href {\doibase
  10.1103/PhysRevLett.86.2050} {\bibfield  {journal} {\bibinfo  {journal}
  {Physical Review Letters}\ }\textbf {\bibinfo {volume} {86}},\ \bibinfo
  {pages} {2050--2053} (\bibinfo {year} {2001})}\BibitemShut {NoStop}%
\bibitem [{\citenamefont {Poulain}\ \emph {et~al.}(2006)\citenamefont
  {Poulain}, \citenamefont {Calvo}, \citenamefont {Antoine}, \citenamefont
  {Broyer},\ and\ \citenamefont {Dugourd}}]{poulain2006performances}%
  \BibitemOpen
  \bibfield  {author} {\bibinfo {author} {\bibfnamefont {P.}~\bibnamefont
  {Poulain}}, \bibinfo {author} {\bibfnamefont {F.}~\bibnamefont {Calvo}},
  \bibinfo {author} {\bibfnamefont {R.}~\bibnamefont {Antoine}}, \bibinfo
  {author} {\bibfnamefont {M.}~\bibnamefont {Broyer}}, \ and\ \bibinfo {author}
  {\bibfnamefont {P.}~\bibnamefont {Dugourd}},\ }\bibfield  {title} {\enquote
  {\bibinfo {title} {Performances of wang-landau algorithms for continuous
  systems},}\ }\href@noop {} {\bibfield  {journal} {\bibinfo  {journal}
  {Physical Review E}\ }\textbf {\bibinfo {volume} {73}},\ \bibinfo {pages}
  {056704} (\bibinfo {year} {2006})}\BibitemShut {NoStop}%
\bibitem [{\citenamefont {Nigra}, \citenamefont {Freeman},\ and\ \citenamefont
  {Doll}(2005)}]{nigra_combining_2005}%
  \BibitemOpen
  \bibfield  {author} {\bibinfo {author} {\bibfnamefont {P.}~\bibnamefont
  {Nigra}}, \bibinfo {author} {\bibfnamefont {D.~L.}\ \bibnamefont {Freeman}},
  \ and\ \bibinfo {author} {\bibfnamefont {J.~D.}\ \bibnamefont {Doll}},\
  }\bibfield  {title} {{\selectlanguage {english}\enquote {\bibinfo {title}
  {Combining smart darting with parallel tempering using {Eckart} space:
  {Application} to {Lennard}–{Jones} clusters},}\ }}\href {\doibase
  10.1063/1.1858433} {\bibfield  {journal} {\bibinfo  {journal} {The Journal of
  Chemical Physics}\ }\textbf {\bibinfo {volume} {122}},\ \bibinfo {pages}
  {114113} (\bibinfo {year} {2005})}\BibitemShut {NoStop}%
\bibitem [{\citenamefont {Sharapov}\ and\ \citenamefont
  {Mandelshtam}(2007)}]{sharapov_solidsolid_2007}%
  \BibitemOpen
  \bibfield  {author} {\bibinfo {author} {\bibfnamefont {V.~A.}\ \bibnamefont
  {Sharapov}}\ and\ \bibinfo {author} {\bibfnamefont {V.~A.}\ \bibnamefont
  {Mandelshtam}},\ }\bibfield  {title} {{\selectlanguage {english}\enquote
  {\bibinfo {title} {{Solid}-{Solid} {Structural} {Transformations} in
  {Lennard}-{Jones} {Clusters}: {Accurate} {Simulations} versus the {Harmonic}
  {Superposition} {Approximation} $^{\textrm{†}}$},}\ }}\href {\doibase
  10.1021/jp072929c} {\bibfield  {journal} {\bibinfo  {journal} {The Journal of
  Physical Chemistry A}\ }\textbf {\bibinfo {volume} {111}},\ \bibinfo {pages}
  {10284--10291} (\bibinfo {year} {2007})}\BibitemShut {NoStop}%
\bibitem [{\citenamefont {Goedecker}(2004)}]{goedecker_minima_2004}%
  \BibitemOpen
  \bibfield  {author} {\bibinfo {author} {\bibfnamefont {S.}~\bibnamefont
  {Goedecker}},\ }\bibfield  {title} {{\selectlanguage {english}\enquote
  {\bibinfo {title} {Minima hopping: {An} efficient search method for the
  global minimum of the potential energy surface of complex molecular
  systems},}\ }}\href {\doibase 10.1063/1.1724816} {\bibfield  {journal}
  {\bibinfo  {journal} {The Journal of Chemical Physics}\ }\textbf {\bibinfo
  {volume} {120}},\ \bibinfo {pages} {9911--9917} (\bibinfo {year}
  {2004})}\BibitemShut {NoStop}%
\bibitem [{\citenamefont {Andricioaei}, \citenamefont {Straub},\ and\
  \citenamefont {Voter}(2001)}]{andricioaei_smart_2001}%
  \BibitemOpen
  \bibfield  {author} {\bibinfo {author} {\bibfnamefont {I.}~\bibnamefont
  {Andricioaei}}, \bibinfo {author} {\bibfnamefont {J.~E.}\ \bibnamefont
  {Straub}}, \ and\ \bibinfo {author} {\bibfnamefont {A.~F.}\ \bibnamefont
  {Voter}},\ }\bibfield  {title} {{\selectlanguage {english}\enquote {\bibinfo
  {title} {Smart {Darting} {Monte} {Carlo}},}\ }}\href {\doibase
  10.1063/1.1358861} {\bibfield  {journal} {\bibinfo  {journal} {The Journal of
  Chemical Physics}\ }\textbf {\bibinfo {volume} {114}},\ \bibinfo {pages}
  {6994--7000} (\bibinfo {year} {2001})}\BibitemShut {NoStop}%
\bibitem [{\citenamefont {Sharapov}, \citenamefont {Meluzzi},\ and\
  \citenamefont {Mandelshtam}(2007)}]{sharapov_low-temperature_2007}%
  \BibitemOpen
  \bibfield  {author} {\bibinfo {author} {\bibfnamefont {V.~A.}\ \bibnamefont
  {Sharapov}}, \bibinfo {author} {\bibfnamefont {D.}~\bibnamefont {Meluzzi}}, \
  and\ \bibinfo {author} {\bibfnamefont {V.~A.}\ \bibnamefont {Mandelshtam}},\
  }\bibfield  {title} {{\selectlanguage {english}\enquote {\bibinfo {title}
  {Low-{Temperature} {Structural} {Transitions}: {Circumventing} the
  {Broken}-{Ergodicity} {Problem}},}\ }}\href {\doibase
  10.1103/PhysRevLett.98.105701} {\bibfield  {journal} {\bibinfo  {journal}
  {Physical Review Letters}\ }\textbf {\bibinfo {volume} {98}} (\bibinfo {year}
  {2007}),\ 10.1103/PhysRevLett.98.105701}\BibitemShut {NoStop}%
\bibitem [{\citenamefont {Uhlherr}\ and\ \citenamefont
  {Theodorou}(2006)}]{minmap}%
  \BibitemOpen
  \bibfield  {author} {\bibinfo {author} {\bibfnamefont {A.}~\bibnamefont
  {Uhlherr}}\ and\ \bibinfo {author} {\bibfnamefont {D.~N.}\ \bibnamefont
  {Theodorou}},\ }\bibfield  {title} {\enquote {\bibinfo {title} {Accelerating
  molecular simulations by reversible mapping between local minima},}\
  }\href@noop {} {\bibfield  {journal} {\bibinfo  {journal} {The Journal of
  chemical physics}\ }\textbf {\bibinfo {volume} {125}},\ \bibinfo {pages}
  {084107} (\bibinfo {year} {2006})}\BibitemShut {NoStop}%
\bibitem [{\citenamefont {Noé}\ \emph {et~al.}(2019)\citenamefont {Noé},
  \citenamefont {Olsson}, \citenamefont {Köhler},\ and\ \citenamefont
  {Wu}}]{noe_boltzmann_2019}%
  \BibitemOpen
  \bibfield  {author} {\bibinfo {author} {\bibfnamefont {F.}~\bibnamefont
  {Noé}}, \bibinfo {author} {\bibfnamefont {S.}~\bibnamefont {Olsson}},
  \bibinfo {author} {\bibfnamefont {J.}~\bibnamefont {Köhler}}, \ and\
  \bibinfo {author} {\bibfnamefont {H.}~\bibnamefont {Wu}},\ }\bibfield
  {title} {{\selectlanguage {english}\enquote {\bibinfo {title} {Boltzmann
  generators: {Sampling} equilibrium states of many-body systems with deep
  learning},}\ }}\href {\doibase 10.1126/science.aaw1147} {\bibfield  {journal}
  {\bibinfo  {journal} {Science}\ }\textbf {\bibinfo {volume} {365}},\ \bibinfo
  {pages} {eaaw1147} (\bibinfo {year} {2019})}\BibitemShut {NoStop}%
\bibitem [{\citenamefont {Bogdan}, \citenamefont {Wales},\ and\ \citenamefont
  {Calvo}(2006)}]{bogdan2006equilibrium}%
  \BibitemOpen
  \bibfield  {author} {\bibinfo {author} {\bibfnamefont {T.~V.}\ \bibnamefont
  {Bogdan}}, \bibinfo {author} {\bibfnamefont {D.~J.}\ \bibnamefont {Wales}}, \
  and\ \bibinfo {author} {\bibfnamefont {F.}~\bibnamefont {Calvo}},\ }\bibfield
   {title} {\enquote {\bibinfo {title} {Equilibrium thermodynamics from
  basin-sampling},}\ }\href@noop {} {\bibfield  {journal} {\bibinfo  {journal}
  {The Journal of chemical physics}\ }\textbf {\bibinfo {volume} {124}},\
  \bibinfo {pages} {044102} (\bibinfo {year} {2006})}\BibitemShut {NoStop}%
\bibitem [{\citenamefont {Wales}(2013)}]{wales_surveying_2013}%
  \BibitemOpen
  \bibfield  {author} {\bibinfo {author} {\bibfnamefont {D.~J.}\ \bibnamefont
  {Wales}},\ }\bibfield  {title} {{\selectlanguage {english}\enquote {\bibinfo
  {title} {Surveying a complex potential energy landscape: {Overcoming} broken
  ergodicity using basin-sampling},}\ }}\href {\doibase
  10.1016/j.cplett.2013.07.066} {\bibfield  {journal} {\bibinfo  {journal}
  {Chemical Physics Letters}\ }\textbf {\bibinfo {volume} {584}},\ \bibinfo
  {pages} {1--9} (\bibinfo {year} {2013})}\BibitemShut {NoStop}%
\bibitem [{\citenamefont {Martiniani}\ \emph {et~al.}(2014)\citenamefont
  {Martiniani}, \citenamefont {Stevenson}, \citenamefont {Wales},\ and\
  \citenamefont {Frenkel}}]{martiniani_superposition_2014}%
  \BibitemOpen
  \bibfield  {author} {\bibinfo {author} {\bibfnamefont {S.}~\bibnamefont
  {Martiniani}}, \bibinfo {author} {\bibfnamefont {J.~D.}\ \bibnamefont
  {Stevenson}}, \bibinfo {author} {\bibfnamefont {D.~J.}\ \bibnamefont
  {Wales}}, \ and\ \bibinfo {author} {\bibfnamefont {D.}~\bibnamefont
  {Frenkel}},\ }\bibfield  {title} {{\selectlanguage {english}\enquote
  {\bibinfo {title} {Superposition {Enhanced} {Nested} {Sampling}},}\ }}\href
  {\doibase 10.1103/PhysRevX.4.031034} {\bibfield  {journal} {\bibinfo
  {journal} {Physical Review X}\ }\textbf {\bibinfo {volume} {4}} (\bibinfo
  {year} {2014}),\ 10.1103/PhysRevX.4.031034}\BibitemShut {NoStop}%
\bibitem [{\citenamefont {Stillinger}(1999)}]{stillinger_exponential_1999}%
  \BibitemOpen
  \bibfield  {author} {\bibinfo {author} {\bibfnamefont {F.~H.}\ \bibnamefont
  {Stillinger}},\ }\bibfield  {title} {{\selectlanguage {english}\enquote
  {\bibinfo {title} {Exponential multiplicity of inherent structures},}\
  }}\href {\doibase 10.1103/PhysRevE.59.48} {\bibfield  {journal} {\bibinfo
  {journal} {Physical Review E}\ }\textbf {\bibinfo {volume} {59}},\ \bibinfo
  {pages} {48--51} (\bibinfo {year} {1999})}\BibitemShut {NoStop}%
\bibitem [{\citenamefont {Doye}, \citenamefont {Miller},\ and\ \citenamefont
  {Wales}(1999{\natexlab{a}})}]{doye_evolution_1999}%
  \BibitemOpen
  \bibfield  {author} {\bibinfo {author} {\bibfnamefont {J.~P.~K.}\
  \bibnamefont {Doye}}, \bibinfo {author} {\bibfnamefont {M.~A.}\ \bibnamefont
  {Miller}}, \ and\ \bibinfo {author} {\bibfnamefont {D.~J.}\ \bibnamefont
  {Wales}},\ }\bibfield  {title} {{\selectlanguage {english}\enquote {\bibinfo
  {title} {Evolution of the potential energy surface with size for
  {Lennard}-{Jones} clusters},}\ }}\href {\doibase 10.1063/1.480217} {\bibfield
   {journal} {\bibinfo  {journal} {The Journal of Chemical Physics}\ }\textbf
  {\bibinfo {volume} {111}},\ \bibinfo {pages} {8417--8428} (\bibinfo {year}
  {1999}{\natexlab{a}})}\BibitemShut {NoStop}%
\bibitem [{\citenamefont {Frantsuzov}\ and\ \citenamefont
  {Mandelshtam}(2005)}]{frantsuzov_size-temperature_2005}%
  \BibitemOpen
  \bibfield  {author} {\bibinfo {author} {\bibfnamefont {P.~A.}\ \bibnamefont
  {Frantsuzov}}\ and\ \bibinfo {author} {\bibfnamefont {V.~A.}\ \bibnamefont
  {Mandelshtam}},\ }\bibfield  {title} {{\selectlanguage {english}\enquote
  {\bibinfo {title} {Size-temperature phase diagram for small {Lennard}-{Jones}
  clusters},}\ }}\href {\doibase 10.1103/PhysRevE.72.037102} {\bibfield
  {journal} {\bibinfo  {journal} {Physical Review E}\ }\textbf {\bibinfo
  {volume} {72}},\ \bibinfo {pages} {037102} (\bibinfo {year}
  {2005})}\BibitemShut {NoStop}%
\bibitem [{\citenamefont {Behler}(2017)}]{behler_first_2017}%
  \BibitemOpen
  \bibfield  {author} {\bibinfo {author} {\bibfnamefont {J.}~\bibnamefont
  {Behler}},\ }\bibfield  {title} {{\selectlanguage {english}\enquote {\bibinfo
  {title} {First {Principles} {Neural} {Network} {Potentials} for {Reactive}
  {Simulations} of {Large} {Molecular} and {Condensed} {Systems}},}\ }}\href
  {\doibase 10.1002/anie.201703114} {\bibfield  {journal} {\bibinfo  {journal}
  {Angewandte Chemie International Edition}\ }\textbf {\bibinfo {volume}
  {56}},\ \bibinfo {pages} {12828--12840} (\bibinfo {year} {2017})}\BibitemShut
  {NoStop}%
\bibitem [{\citenamefont {Eckart}(1935)}]{eckart_studies_1935}%
  \BibitemOpen
  \bibfield  {author} {\bibinfo {author} {\bibfnamefont {C.}~\bibnamefont
  {Eckart}},\ }\bibfield  {title} {{\selectlanguage {english}\enquote {\bibinfo
  {title} {Some {Studies} {Concerning} {Rotating} {Axes} and {Polyatomic}
  {Molecules}},}\ }}\href {\doibase 10.1103/PhysRev.47.552} {\bibfield
  {journal} {\bibinfo  {journal} {Physical Review}\ }\textbf {\bibinfo {volume}
  {47}},\ \bibinfo {pages} {552--558} (\bibinfo {year} {1935})}\BibitemShut
  {NoStop}%
\bibitem [{\citenamefont {Kudin}\ and\ \citenamefont
  {Dymarsky}(2005)}]{kudin_eckart_2005}%
  \BibitemOpen
  \bibfield  {author} {\bibinfo {author} {\bibfnamefont {K.~N.}\ \bibnamefont
  {Kudin}}\ and\ \bibinfo {author} {\bibfnamefont {A.~Y.}\ \bibnamefont
  {Dymarsky}},\ }\bibfield  {title} {{\selectlanguage {english}\enquote
  {\bibinfo {title} {Eckart axis conditions and the minimization of the
  root-mean-square deviation: {Two} closely related problems},}\ }}\href
  {\doibase 10.1063/1.1929739} {\bibfield  {journal} {\bibinfo  {journal} {The
  Journal of Chemical Physics}\ }\textbf {\bibinfo {volume} {122}},\ \bibinfo
  {pages} {224105} (\bibinfo {year} {2005})}\BibitemShut {NoStop}%
\bibitem [{\citenamefont {Kearsley}(1989)}]{Kearsley_1989}%
  \BibitemOpen
  \bibfield  {author} {\bibinfo {author} {\bibfnamefont {S.~K.}\ \bibnamefont
  {Kearsley}},\ }\bibfield  {title} {\enquote {\bibinfo {title} {{On the
  orthogonal transformation used for structural comparisons}},}\ }\href
  {\doibase 10.1107/S0108767388010128} {\bibfield  {journal} {\bibinfo
  {journal} {Acta Crystallographica Section A}\ }\textbf {\bibinfo {volume}
  {45}},\ \bibinfo {pages} {208--210} (\bibinfo {year} {1989})}\BibitemShut
  {NoStop}%
\bibitem [{\citenamefont {Coutsias}, \citenamefont {Seok},\ and\ \citenamefont
  {Dill}()}]{Coutsias_Using_2004}%
  \BibitemOpen
  \bibfield  {author} {\bibinfo {author} {\bibfnamefont {E.~A.}\ \bibnamefont
  {Coutsias}}, \bibinfo {author} {\bibfnamefont {C.}~\bibnamefont {Seok}}, \
  and\ \bibinfo {author} {\bibfnamefont {K.~A.}\ \bibnamefont {Dill}},\
  }\bibfield  {title} {\enquote {\bibinfo {title} {Using quaternions to
  calculate rmsd},}\ }\href {\doibase 10.1002/jcc.20110} {\bibfield  {journal}
  {\bibinfo  {journal} {Journal of Computational Chemistry}\ }\textbf {\bibinfo
  {volume} {25}},\ \bibinfo {pages} {1849--1857}}\BibitemShut {NoStop}%
\bibitem [{\citenamefont {Krasnoshchekov}, \citenamefont {Isayeva},\ and\
  \citenamefont {Stepanov}(2014)}]{krasnoshchekov_determination_2014}%
  \BibitemOpen
  \bibfield  {author} {\bibinfo {author} {\bibfnamefont {S.~V.}\ \bibnamefont
  {Krasnoshchekov}}, \bibinfo {author} {\bibfnamefont {E.~V.}\ \bibnamefont
  {Isayeva}}, \ and\ \bibinfo {author} {\bibfnamefont {N.~F.}\ \bibnamefont
  {Stepanov}},\ }\bibfield  {title} {{\selectlanguage {english}\enquote
  {\bibinfo {title} {Determination of the {Eckart} molecule-fixed frame by use
  of the apparatus of quaternion algebra},}\ }}\href {\doibase
  10.1063/1.4870936} {\bibfield  {journal} {\bibinfo  {journal} {The Journal of
  Chemical Physics}\ }\textbf {\bibinfo {volume} {140}},\ \bibinfo {pages}
  {154104} (\bibinfo {year} {2014})}\BibitemShut {NoStop}%
\bibitem [{\citenamefont {Kuhn}(1955)}]{kuhn_hungarian_1955}%
  \BibitemOpen
  \bibfield  {author} {\bibinfo {author} {\bibfnamefont {H.~W.}\ \bibnamefont
  {Kuhn}},\ }\bibfield  {title} {{\selectlanguage {english}\enquote {\bibinfo
  {title} {The {Hungarian} method for the assignment problem},}\ }}\href
  {\doibase 10.1002/nav.3800020109} {\bibfield  {journal} {\bibinfo  {journal}
  {Naval Research Logistics Quarterly}\ }\textbf {\bibinfo {volume} {2}},\
  \bibinfo {pages} {83--97} (\bibinfo {year} {1955})}\BibitemShut {NoStop}%
\bibitem [{\citenamefont {Jonker}\ and\ \citenamefont
  {Volgenant}(1987)}]{jonker1987shortest}%
  \BibitemOpen
  \bibfield  {author} {\bibinfo {author} {\bibfnamefont {R.}~\bibnamefont
  {Jonker}}\ and\ \bibinfo {author} {\bibfnamefont {A.}~\bibnamefont
  {Volgenant}},\ }\bibfield  {title} {\enquote {\bibinfo {title} {A shortest
  augmenting path algorithm for dense and sparse linear assignment problems},}\
  }\href@noop {} {\bibfield  {journal} {\bibinfo  {journal} {Computing}\
  }\textbf {\bibinfo {volume} {38}},\ \bibinfo {pages} {325--340} (\bibinfo
  {year} {1987})}\BibitemShut {NoStop}%
\bibitem [{\citenamefont {Schaefer}\ and\ \citenamefont
  {Goedecker}(2016)}]{schaefer_computationally_2016}%
  \BibitemOpen
  \bibfield  {author} {\bibinfo {author} {\bibfnamefont {B.}~\bibnamefont
  {Schaefer}}\ and\ \bibinfo {author} {\bibfnamefont {S.}~\bibnamefont
  {Goedecker}},\ }\bibfield  {title} {{\selectlanguage {english}\enquote
  {\bibinfo {title} {Computationally efficient characterization of potential
  energy surfaces based on fingerprint distances},}\ }}\href {\doibase
  10.1063/1.4956461} {\bibfield  {journal} {\bibinfo  {journal} {The Journal of
  Chemical Physics}\ }\textbf {\bibinfo {volume} {145}},\ \bibinfo {pages}
  {034101} (\bibinfo {year} {2016})}\BibitemShut {NoStop}%
\bibitem [{\citenamefont {Gentle}(2009)}]{gentle_computational_2009}%
  \BibitemOpen
  \bibfield  {author} {\bibinfo {author} {\bibfnamefont {J.~E.}\ \bibnamefont
  {Gentle}},\ }\href@noop {} {\emph {\bibinfo {title} {Computational
  statistics}}},\ Statistics and computing\ (\bibinfo  {publisher} {Springer},\
  \bibinfo {address} {Dordrecht ; New York},\ \bibinfo {year} {2009})\ pp.\
  \bibinfo {pages} {315--316},\ \bibinfo {note} {oCLC:
  ocn437081973}\BibitemShut {NoStop}%
\bibitem [{\citenamefont {Dempster}, \citenamefont {Laird},\ and\ \citenamefont
  {Rubin}(1977)}]{dempster_maximum_1977}%
  \BibitemOpen
  \bibfield  {author} {\bibinfo {author} {\bibfnamefont {A.~P.}\ \bibnamefont
  {Dempster}}, \bibinfo {author} {\bibfnamefont {N.~M.}\ \bibnamefont {Laird}},
  \ and\ \bibinfo {author} {\bibfnamefont {D.~B.}\ \bibnamefont {Rubin}},\
  }\bibfield  {title} {{\selectlanguage {english}\enquote {\bibinfo {title}
  {Maximum {Likelihood} from {Incomplete} {Data} via the {EM} {Algorithm}},}\
  }}\href@noop {} {\bibfield  {journal} {\bibinfo  {journal} {Journal of the
  Royal Statistical Society. Series B (Methodological)}\ }\textbf {\bibinfo
  {volume} {39}},\ \bibinfo {pages} {1--38} (\bibinfo {year}
  {1977})}\BibitemShut {NoStop}%
\bibitem [{\citenamefont {Bilmes}(1998)}]{bilmes_1998}%
  \BibitemOpen
  \bibfield  {author} {\bibinfo {author} {\bibfnamefont {J.}~\bibnamefont
  {Bilmes}},\ }\href
  {http://citeseerx.ist.psu.edu/viewdoc/summary?doi=10.1.1.28.613} {\enquote
  {\bibinfo {title} {A gentle tutorial of the em algorithm and its application
  to parameter estimation for gaussian mixture and hidden markov models},}\
  }\bibinfo {type} {Tech. Rep.}\ (\bibinfo {year} {1998})\BibitemShut {NoStop}%
\bibitem [{\citenamefont {Gupta}\ and\ \citenamefont
  {Chen}(2010)}]{gupta_2010}%
  \BibitemOpen
  \bibfield  {author} {\bibinfo {author} {\bibfnamefont {M.~R.}\ \bibnamefont
  {Gupta}}\ and\ \bibinfo {author} {\bibfnamefont {Y.}~\bibnamefont {Chen}},\
  }\bibfield  {title} {\enquote {\bibinfo {title} {Theory and use of the em
  algorithm.}}\ }\href
  {http://dblp.uni-trier.de/db/journals/ftsig/ftsig4.html#GuptaC10} {\bibfield
  {journal} {\bibinfo  {journal} {Foundations and Trends in Signal Processing}\
  }\textbf {\bibinfo {volume} {4}},\ \bibinfo {pages} {223--296} (\bibinfo
  {year} {2010})}\BibitemShut {NoStop}%
\bibitem [{\citenamefont {Brooks}, \citenamefont {Jane{\v{z}}i{\v{c}}},\ and\
  \citenamefont {Karplus}(1995)}]{brooks1995harmonic}%
  \BibitemOpen
  \bibfield  {author} {\bibinfo {author} {\bibfnamefont {B.~R.}\ \bibnamefont
  {Brooks}}, \bibinfo {author} {\bibfnamefont {D.}~\bibnamefont
  {Jane{\v{z}}i{\v{c}}}}, \ and\ \bibinfo {author} {\bibfnamefont
  {M.}~\bibnamefont {Karplus}},\ }\bibfield  {title} {\enquote {\bibinfo
  {title} {Harmonic analysis of large systems. i. methodology},}\ }\href@noop
  {} {\bibfield  {journal} {\bibinfo  {journal} {Journal of computational
  chemistry}\ }\textbf {\bibinfo {volume} {16}},\ \bibinfo {pages} {1522--1542}
  (\bibinfo {year} {1995})}\BibitemShut {NoStop}%
\bibitem [{\citenamefont {Koehler}(1966)}]{koehler1966theory}%
  \BibitemOpen
  \bibfield  {author} {\bibinfo {author} {\bibfnamefont {T.~R.}\ \bibnamefont
  {Koehler}},\ }\bibfield  {title} {\enquote {\bibinfo {title} {Theory of the
  self-consistent harmonic approximation with application to solid neon},}\
  }\href@noop {} {\bibfield  {journal} {\bibinfo  {journal} {Physical Review
  Letters}\ }\textbf {\bibinfo {volume} {17}},\ \bibinfo {pages} {89} (\bibinfo
  {year} {1966})}\BibitemShut {NoStop}%
\bibitem [{\citenamefont {Gillis}, \citenamefont {Werthamer},\ and\
  \citenamefont {Koehler}(1968)}]{gillis1968properties}%
  \BibitemOpen
  \bibfield  {author} {\bibinfo {author} {\bibfnamefont {N.}~\bibnamefont
  {Gillis}}, \bibinfo {author} {\bibfnamefont {N.}~\bibnamefont {Werthamer}}, \
  and\ \bibinfo {author} {\bibfnamefont {T.}~\bibnamefont {Koehler}},\
  }\bibfield  {title} {\enquote {\bibinfo {title} {Properties of crystalline
  argon and neon in the self-consistent phonon approximation},}\ }\href@noop {}
  {\bibfield  {journal} {\bibinfo  {journal} {Physical Review}\ }\textbf
  {\bibinfo {volume} {165}},\ \bibinfo {pages} {951} (\bibinfo {year}
  {1968})}\BibitemShut {NoStop}%
\bibitem [{\citenamefont {Georgescu}\ and\ \citenamefont
  {Mandelshtam}(2012)}]{georgescu2012self}%
  \BibitemOpen
  \bibfield  {author} {\bibinfo {author} {\bibfnamefont {I.}~\bibnamefont
  {Georgescu}}\ and\ \bibinfo {author} {\bibfnamefont {V.~A.}\ \bibnamefont
  {Mandelshtam}},\ }\bibfield  {title} {\enquote {\bibinfo {title}
  {Self-consistent phonons revisited. i. the role of thermal versus quantum
  fluctuations on structural transitions in large lennard-jones clusters},}\
  }\href@noop {} {\bibfield  {journal} {\bibinfo  {journal} {The Journal of
  chemical physics}\ }\textbf {\bibinfo {volume} {137}},\ \bibinfo {pages}
  {144106} (\bibinfo {year} {2012})}\BibitemShut {NoStop}%
\bibitem [{\citenamefont {Duane}\ \emph {et~al.}(1987)\citenamefont {Duane},
  \citenamefont {Kennedy}, \citenamefont {Pendleton},\ and\ \citenamefont
  {Roweth}}]{duane_1987}%
  \BibitemOpen
  \bibfield  {author} {\bibinfo {author} {\bibfnamefont {S.}~\bibnamefont
  {Duane}}, \bibinfo {author} {\bibfnamefont {A.~D.}\ \bibnamefont {Kennedy}},
  \bibinfo {author} {\bibfnamefont {B.~J.}\ \bibnamefont {Pendleton}}, \ and\
  \bibinfo {author} {\bibfnamefont {D.}~\bibnamefont {Roweth}},\ }\bibfield
  {title} {\enquote {\bibinfo {title} {Hybrid monte carlo},}\ }\href {\doibase
  DOI: 10.1016/0370-2693(87)91197-X} {\bibfield  {journal} {\bibinfo  {journal}
  {Physics Letters B}\ }\textbf {\bibinfo {volume} {195}},\ \bibinfo {pages}
  {216 -- 222} (\bibinfo {year} {1987})}\BibitemShut {NoStop}%
\bibitem [{\citenamefont {Neal}(2012)}]{neal_mcmc_2012}%
  \BibitemOpen
  \bibfield  {author} {\bibinfo {author} {\bibfnamefont {R.~M.}\ \bibnamefont
  {Neal}},\ }\bibfield  {title} {{\selectlanguage {english}\enquote {\bibinfo
  {title} {{MCMC} using {Hamiltonian} dynamics},}\ }}\href
  {http://arxiv.org/abs/1206.1901} {\bibfield  {journal} {\bibinfo  {journal}
  {arXiv:1206.1901 [physics, stat]}\ } (\bibinfo {year} {2012})},\ \bibinfo
  {note} {arXiv: 1206.1901}\BibitemShut {NoStop}%
\bibitem [{\citenamefont {Swendsen}\ and\ \citenamefont
  {Wang}(1986)}]{swendsen_replica_1986}%
  \BibitemOpen
  \bibfield  {author} {\bibinfo {author} {\bibfnamefont {R.~H.}\ \bibnamefont
  {Swendsen}}\ and\ \bibinfo {author} {\bibfnamefont {J.-S.}\ \bibnamefont
  {Wang}},\ }\bibfield  {title} {{\selectlanguage {english}\enquote {\bibinfo
  {title} {Replica {Monte} {Carlo} {Simulation} of {Spin}-{Glasses}},}\ }}\href
  {\doibase 10.1103/PhysRevLett.57.2607} {\bibfield  {journal} {\bibinfo
  {journal} {Physical Review Letters}\ }\textbf {\bibinfo {volume} {57}},\
  \bibinfo {pages} {2607--2609} (\bibinfo {year} {1986})}\BibitemShut {NoStop}%
\bibitem [{\citenamefont {Neirotti}\ \emph {et~al.}(2000)\citenamefont
  {Neirotti}, \citenamefont {Calvo}, \citenamefont {Freeman},\ and\
  \citenamefont {Doll}}]{neirotti_phase_2000}%
  \BibitemOpen
  \bibfield  {author} {\bibinfo {author} {\bibfnamefont {J.~P.}\ \bibnamefont
  {Neirotti}}, \bibinfo {author} {\bibfnamefont {F.}~\bibnamefont {Calvo}},
  \bibinfo {author} {\bibfnamefont {D.~L.}\ \bibnamefont {Freeman}}, \ and\
  \bibinfo {author} {\bibfnamefont {J.~D.}\ \bibnamefont {Doll}},\ }\bibfield
  {title} {{\selectlanguage {english}\enquote {\bibinfo {title} {Phase changes
  in 38-atom {Lennard}-{Jones} clusters. {I}. {A} parallel tempering study in
  the canonical ensemble},}\ }}\href {\doibase 10.1063/1.481671} {\bibfield
  {journal} {\bibinfo  {journal} {The Journal of Chemical Physics}\ }\textbf
  {\bibinfo {volume} {112}},\ \bibinfo {pages} {10340--10349} (\bibinfo {year}
  {2000})}\BibitemShut {NoStop}%
\bibitem [{\citenamefont {Earl}\ and\ \citenamefont
  {Deem}(2005)}]{earl_parallel_2005}%
  \BibitemOpen
  \bibfield  {author} {\bibinfo {author} {\bibfnamefont {D.~J.}\ \bibnamefont
  {Earl}}\ and\ \bibinfo {author} {\bibfnamefont {M.~W.}\ \bibnamefont
  {Deem}},\ }\bibfield  {title} {{\selectlanguage {english}\enquote {\bibinfo
  {title} {Parallel tempering: {Theory}, applications, and new perspectives},}\
  }}\href {\doibase 10.1039/b509983h} {\bibfield  {journal} {\bibinfo
  {journal} {Physical Chemistry Chemical Physics}\ }\textbf {\bibinfo {volume}
  {7}},\ \bibinfo {pages} {3910} (\bibinfo {year} {2005})}\BibitemShut
  {NoStop}%
\bibitem [{\citenamefont {Kofke}(2002)}]{kofke_acceptance_2002}%
  \BibitemOpen
  \bibfield  {author} {\bibinfo {author} {\bibfnamefont {D.~A.}\ \bibnamefont
  {Kofke}},\ }\bibfield  {title} {{\selectlanguage {english}\enquote {\bibinfo
  {title} {On the acceptance probability of replica-exchange {Monte} {Carlo}
  trials},}\ }}\href {\doibase 10.1063/1.1507776} {\bibfield  {journal}
  {\bibinfo  {journal} {The Journal of Chemical Physics}\ }\textbf {\bibinfo
  {volume} {117}},\ \bibinfo {pages} {6911--6914} (\bibinfo {year}
  {2002})}\BibitemShut {NoStop}%
\bibitem [{\citenamefont {Rathore}, \citenamefont {Chopra},\ and\ \citenamefont
  {de~Pablo}(2005)}]{rathore_optimal_2005}%
  \BibitemOpen
  \bibfield  {author} {\bibinfo {author} {\bibfnamefont {N.}~\bibnamefont
  {Rathore}}, \bibinfo {author} {\bibfnamefont {M.}~\bibnamefont {Chopra}}, \
  and\ \bibinfo {author} {\bibfnamefont {J.~J.}\ \bibnamefont {de~Pablo}},\
  }\bibfield  {title} {{\selectlanguage {english}\enquote {\bibinfo {title}
  {Optimal allocation of replicas in parallel tempering simulations},}\ }}\href
  {\doibase 10.1063/1.1831273} {\bibfield  {journal} {\bibinfo  {journal} {The
  Journal of Chemical Physics}\ }\textbf {\bibinfo {volume} {122}},\ \bibinfo
  {pages} {024111} (\bibinfo {year} {2005})}\BibitemShut {NoStop}%
\bibitem [{\citenamefont {Doye}, \citenamefont {Miller},\ and\ \citenamefont
  {Wales}(1999{\natexlab{b}})}]{doye_double-funnel_1999}%
  \BibitemOpen
  \bibfield  {author} {\bibinfo {author} {\bibfnamefont {J.~P.~K.}\
  \bibnamefont {Doye}}, \bibinfo {author} {\bibfnamefont {M.~A.}\ \bibnamefont
  {Miller}}, \ and\ \bibinfo {author} {\bibfnamefont {D.~J.}\ \bibnamefont
  {Wales}},\ }\bibfield  {title} {{\selectlanguage {english}\enquote {\bibinfo
  {title} {The double-funnel energy landscape of the 38-atom {Lennard}-{Jones}
  cluster},}\ }}\href {\doibase 10.1063/1.478595} {\bibfield  {journal}
  {\bibinfo  {journal} {The Journal of Chemical Physics}\ }\textbf {\bibinfo
  {volume} {110}},\ \bibinfo {pages} {6896--6906} (\bibinfo {year}
  {1999}{\natexlab{b}})}\BibitemShut {NoStop}%
\bibitem [{\citenamefont {Johansson}\ and\ \citenamefont
  {Veryazov}(2017)}]{johansson_automatic_2017}%
  \BibitemOpen
  \bibfield  {author} {\bibinfo {author} {\bibfnamefont {M.}~\bibnamefont
  {Johansson}}\ and\ \bibinfo {author} {\bibfnamefont {V.}~\bibnamefont
  {Veryazov}},\ }\bibfield  {title} {{\selectlanguage {english}\enquote
  {\bibinfo {title} {Automatic procedure for generating symmetry adapted
  wavefunctions},}\ }}\href {\doibase 10.1186/s13321-017-0193-3} {\bibfield
  {journal} {\bibinfo  {journal} {Journal of Cheminformatics}\ }\textbf
  {\bibinfo {volume} {9}},\ \bibinfo {pages} {8} (\bibinfo {year}
  {2017})}\BibitemShut {NoStop}%
\end{thebibliography}%

\end{document}